\documentclass{article}
\usepackage{wrapfig}
\usepackage{nth}
\usepackage{multirow}
\usepackage{arxiv}
\usepackage{graphicx}
\usepackage{subcaption} 
\usepackage{float}
\usepackage{longtable}

\usepackage[utf8]{inputenc} 
\usepackage[T1]{fontenc}    
\usepackage{hyperref}       
\usepackage{url}            
\usepackage{booktabs}       
\usepackage{amsfonts}       
\usepackage{nicefrac}       
\usepackage{microtype}      
\usepackage{lipsum}
\usepackage{graphicx}
\graphicspath{ {./images/} }

\title{Gaze Prediction as a Function of Eye Movement Type and Individual Differences}

\author{
 Kateryna Melnyk \thanks{corresponding author} \\
  Texas State University\\
  San Marcos, Texas, 78640, USA \\
  \texttt{k\_m825@txstate.edu} \\
   \And
   Lee Friedman \\
  Texas State University\\
  San Marcos, Texas, 78640, USA \\
  \texttt{l\_f96@txstate.edu} \\
  \And
 Dmytro Katrychuk \\
  Texas State University\\
  San Marcos, Texas, 78640, USA \\
  \texttt{d\_k139@txstate.edu} \\
  \And
 Oleg Komogortsev \\
  Texas State University\\
  San Marcos, Texas, 78640, USA \\
  \texttt{ok@txstate.edu} \\
}

\begin{document}

\title{Gaze Prediction as a Function of Eye Movement Type and Individual Differences}
\maketitle

\begin{abstract}
Eye movement prediction is a promising area of research with the potential to improve performance and the user experience of systems based on eye-tracking technology. In this study, we analyze individual differences in gaze prediction performance. We use three fundamentally different models within the analysis: the lightweight Long Short-Term Memory network (LSTM), the transformer-based network for multivariate time series representation learning (TST), and the Oculomotor Plant Mathematical Model wrapped in the Kalman Filter framework (OPKF). Each solution was assessed on different eye-movement types. We show important subject-to-subject variation for all models and eye-movement types. We found that fixation noise is associated with poorer gaze prediction in fixation. For saccades, higher velocities are associated with poorer gaze prediction performance. We think these individual differences are important and propose that future research should report statistics related to inter-subject variation. We also propose that future models should be designed to reduce subject-to-subject variation.
\end{abstract}


\section{Introduction} Millions of people are excited by the opportunity to participate in various entertainment activities, build new social connections, or experience immersive surroundings. The consumer tier of Extended Reality (XR) devices is growing, and they are in wide usage in training \cite{xie2021review}, healthcare\cite{makinen2022user}, and education \cite{freina2015literature}. Of course, XR technologies operate under resource constraints. One of the main challenges is to improve visual fidelity without increasing the rendering cost in graphically demanding workloads \cite{matthews2020rendering}.

The eye has a high visual central acuity region known as the fovea. Visual acuity decreases towards the periphery, first steeply and then more slowly. One optimization technique that utilizes this varying acuity across the retina is Foveated Rendering (FR). It renders the image at a lower resolution in the peripheral area but at a high resolution where the foveal vision is. FR requires an eye-tracking (ET) system to track real-time eye movement data at a high speed. FR is sensitive to the latency introduced by the ET system. If the system delay is high, the FR might render a high-resolution image for a point that is no longer viewed by the fovea. The acceptable value of system latency is below 40 ms \cite{albert2017latency, stauffert2020latency}, which is necessary for real-time processing. Most current head-mounted displays (HMDs) with eye-tracking cannot meet this criteria \cite{stein2021comparison}. Eye movement prediction helps make FR more efficient and reduces latency by pre-rendering the content within the predicted future gaze area.

Researchers employ various strategies for gaze prediction to design efficient solutions that can operate in real-time. The different devices produce various data types for potential gaze prediction, including eye position, head movement, and saliency maps \cite{hu2020dgaze, hu2021fixationnet, illahi2022real}. The validation and evaluation phases are crucial in any eye movement prediction framework, as eye-tracking data is naturally susceptible to various forms of noise introduced by participants and hardware and recording conditions \cite{holmqvist2022small}. The main focus of the current work is to examine how model prediction performance varies from subject to subject. We present a more in-depth analysis of individual performance differences that might be employed to improve the eye movement prediction approaches. Our study addresses a critical gap in the existing literature by proposing and showing how to evaluate the model's performance on a per-subject basis, as any solution should perform equally well for each individual to ensure that it accurately reflects personal variations in eye movement patterns. In the main report, all results employ a 40 ms PI. Data for other PIs are included in the Appendix.
\vspace{-3mm}

\section{Prior Work}
\subsection{Eye movement Prediction: Approaches}
There has been increasing academic interest in eye movement prediction. At first, researchers concentrated on identifying the areas or objects likely to attract the user's attention in naturalistic scenarios. These models learned subject-dependent saliency maps and predicted fixation regions \cite{chong2018connecting, cornia2018sam}. Along with attention prediction, other groups focused on predicting the landing points of saccades. Different methods were utilized for making predictions: (1) mathematical approaches based on ballistic characteristics of saccades \cite{paeye2016visual, arabadzhiyska2017saccade}; (2) anatomically inspired Oculomotor Plant Mathematical Models (OPMMs) \cite{komogortsev2008eye, komogortsev2009eye}; (3) deep learning approaches using Recurrent Neural Networks (RNNs) \cite{morales2018saccade}; and (5) Long Short-Term Memory Networks (LSTM) \cite{griffith2020shift, morales2021saccade}. 

As eye-tracking technologies improved, studies focused on gaze prediction of the entire position signal. Success is a function of the format of the data and the time interval over which the gaze prediction is made. In several studies \cite{rolff2022gazetransformer, hu2021fixationnet, hu2020dgaze}, the prediction interval (PI) value was more than 150 ms. The average latency for VR applications is under 30 ms. This makes shorter PIs of 20 to 60 ms more relevant.

Hu et al. (\cite{hu2019sgaze, hu2020dgaze, hu2020gaze, hu2021fixationnet}) evaluated gaze prediction with virtual reality (VR) data. Each study provided a detailed correlation analysis of participants' gaze and head movement data alongside VR graphical content information such as task-related data, saliency maps, and object positions in dynamic scenes. In \cite{hu2020dgaze}, the authors applied a Convolutional Neural Network (CNN) based solution that combines object positions in VR scenes, saliency maps, head velocity data, and, if available, users' gaze positions (DGaze-ET) to predict future gaze data. In a subsequent study \cite{hu2021fixationnet}, the FixationNet model was introduced to forecast human eye fixations. Another research group \cite{illahi2022real} approached eye movement prediction with LSTM-based architecture using HMD head rotation and past gaze data.

The transformer model was also applied to gaze prediction and was widely appreciated as a successful approach \cite{wen2022transformers}. Mazzeo et al. \cite{mazzeo2021deep} tested several CNN architectures to find the most accurate gaze estimation solution. Later, they used estimated gaze vectors to predict gaze locations with a combination of LSTM and Transformer architectures based on a self-attention mechanism. In \cite{rolff2022gazetransformer}, the authors experimented with the transformer-based model (GazeTransformer) that used gaze and head position data as input. GazeTransformer was assessed using angular error over a 150 ms PI. 

Some research groups have also conceived of the eye movement prediction task as a sequential decision-making problem. Several works established Reinforcement Learning (RL) solutions that enhance human attention localization \cite{baee2021medirl, yang2020predicting, lv2020improving}. In \cite{lv2020improving}, researchers showed the effectiveness of the RL approach as the driver for the gaze prediction domain, whereas in the \cite{baee2021medirl} study, the maximum entropy deep inverse RL model was demonstrated to be effective in rear-end driving collision scenarios. Moreover, \cite{transformer_based} introduces a transformer-based RL solution for predicting human gaze behavior during video viewing.
\vspace{-2mm}

\subsection{Eye movement Prediction: Evaluation and Metrics}
Several metrics and visualization techniques have been used for eye movement prediction. In most studies \cite{illahi2022real, hu2020dgaze, rolff2022gazetransformer}, the prediction error is calculated as the angular distance between the ground truth and the predicted gaze position expressed in degrees of visual angle (dva). The cumulative distribution function (CDF) plots, with the x-axis representing prediction errors across the entire recording and the y-axis displaying the proportion of data, has become a standard method for assessing performance. 

Rolff et al. \cite{rolff2022metrics} introduced several valuable metrics such as model consistency, partitioning error, overestimation, and underestimation rate for the time-to-event prediction. They propose that model evaluation should be done over different parts of time-to-saccade sequences to capture prediction behavior. Of course, the aggregated prediction error over the entire recording may only partially represent the actual predictive performance of the model. An eye-tracking signal consists of several events, including fixations, saccades, smooth pursuits, blinks, and microsaccades. Each eye-movement type will undoubtedly have unique characteristics, and some eye-movement types may be more or less challenging for the prediction algorithms. In the \cite{illahi2022real} study, researchers illustrate the prediction errors at different normalized gaze velocities. They found that high-velocity events (i.e., saccades) were more challenging for gaze prediction.

Aziz et al \cite{aziz2023practical} propose that prediction should focus on the time period immediately after saccade endings. This period is referred to as the Critical Evaluation Period (CEP). The metric creation was influenced by a phenomenon known as saccadic suppression, during which visual information is not processed \cite{diamond2000ext}. The CEP was defined as the 100 ms interval post-saccade. The success of CEP interval prediction was a function of the amplitude of the prior saccade, with large saccades presenting more error. Another issue is the role of signal quality on gaze prediction \cite{illahi2022real}. These authors assessed model performance on degraded signals and showed that for their approach, the model could successfully deal with lower-quality signals.
\vspace{-2mm}

\section{Methodology}

\subsection{Dataset}
We employed the ``Gazebase Dataset'' for our research. It has already been used in several gaze prediction studies \cite{aziz2023practical, illahi2022real}. Refer to ``GazeBase'' report \cite{griffith2021gazebase} for a detailed overview of the participants, tasks, stimuli descriptions, and all relevant details. All the recruited participants were undergraduate students from Texas State University. The dataset consists of multiple 'Rounds' of data collection. We analyzed only Round 1, which has 322 participants. In this study, we used Session 1 of the random saccades task (RAN). The target was displaced at random locations across the monitor to induce visually guided oblique saccades with varying amplitudes. The displacements of the target ranged from $\pm15^o$ to $\pm9^o$ in the horizontal and vertical directions.

\textit{Ethics and Privacy Statement:} All participants provided informed consent following a protocol approved by the Institutional Research Board (IRB) at Texas State University prior to each round of recording. As part of the consent process, participants acknowledged that their data may be shared in a de-identified form for research purposes. No participant-identifying information is included in the ``GazeBase Dataset''.

\subsection{Prediction Models}
\subsubsection{LSTM}
As noted above, one of the most widely used deep learning architectures for the eye movement prediction problem was the LSTM model \cite{mazzeo2021deep, morales2021saccade, illahi2022real, aziz2023practical}. In the current study, we opted for a lightweight implementation consisting of two recurrent LSTM and two fully connected layers. The LSTM part had a hidden size of 32. The first fully connected layer consists of 32 nodes and the second one has 16 nodes. For a detailed description, see \cite{aziz2023practical}.
\vspace{-2mm}

\subsubsection{TST}
We also used the transformer-based architecture TSTPlus \cite{TST_network}. We selected the Tsai library for implementation \cite{tsai} as it was created specifically for time series tasks. The TST model consists of multiple encoder layers, and each encoder layer incorporates multi-head self-attention mechanisms. This design allows the model to capture temporal patterns by weighing the importance of different time steps in the input sequence. The custom head used for this model has the same structure as the LSTM solution, consisting of two fully connected layers, with the first layer followed by a ReLU activation function. For both models, we performed hyperparameter tuning and experimented with the model parameters.
\vspace{-2mm}

\subsubsection{OPKF}
One of the most widely adopted anatomically inspired OPMMs was introduced in 1980 \cite{bahill1980development}. Since then, the model has established its practical use in various applications, including saccade simulation tailored on a per-subject level \cite{melnyk2024per, katrychuk2022study}, eye movement classification \cite{wadehn2018estimation}, biometrics research \cite{komogortsev2010biom, komogortsev2012biom}, and gaze prediction \cite{komogortsev2008eye, komogortsev2009eye}. We included predictions made with the OPMM wrapped in the Kalman Filter framework (OPKF) \cite{katrychuk2022study} for gaze prediction.

A 13-parameter OPMM model was used with a slight modification made to the neural pulse calculation. The model parameters were estimated using the per-subject optimization procedure outlined in \cite{katrychuk2022study}. The Nelder-Mead optimization method was employed. The OPKF predicts an \textit{a priori} estimate using the current state and a mathematical model. It then updates this prediction with the next available measurement (gaze position and velocity) for the observed part of the state. In simpler terms, at each time-step $t$, the OPKF receives an input consisting of the eye movement classification and a state vector $x^t$, which includes the gaze position and velocity. It then outputs the predicted future gaze vector for time $x^{t + PI}$ where $PI$ - is the prediction interval.
\vspace{-2mm}

\subsection{Data Pre-processing}
For the OPKF prediction framework, we needed eye-tracking signals to be classified into eye movement events. An improved version of the MNH event classification algorithm \cite{mnh} was employed. To execute a detailed analysis of gaze prediction as a function of eye movement type or as a function of individual differences, a set of eye movement characteristics \cite{melnyk2024can, rigas2018study} and signal quality measures \cite{raju2024temp, lohr2019eval}, such as spatial accuracy and precision, were calculated for each recording.

Each recording consists of horizontal and vertical gaze positions, measured in dva. For deep learning models, velocity data was used as input, and the displacement between gaze positions was used as output. For the OPKF model, gaze positional and velocity data, along with eye movement classification labels, are required. The model output is the predicted gaze position.

\subsection{Training}
The subjects were randomly split between the train and test datasets to reduce the risk of overfitting. We ended up with 255 subjects in the training dataset and 67 in the test dataset. The sliding window approach was used to train deep learning models, and the window length was set to 100 ms. The windows were shuffled before splitting into batches of size 256. To facilitate a smoother training process, the PyTorch Lightning framework \cite{falcon2019pytorch} was applied. The loss function was defined as the mean Euclidean distance between the ground truth and predicted gaze data, and we utilized the Adam optimizer with a learning rate of $0.0003$. The code and data required to replicate the results will be made publicly available at the following link: [Link will be provided in future].

\vspace{-2mm}
\section{Results}

\begin{figure*}
    \centering
    \caption{CDF Plots Across All Eye-Movement Events}
    \begin{subfigure}{0.31\textwidth}
        \includegraphics[width=\textwidth, height=4cm]{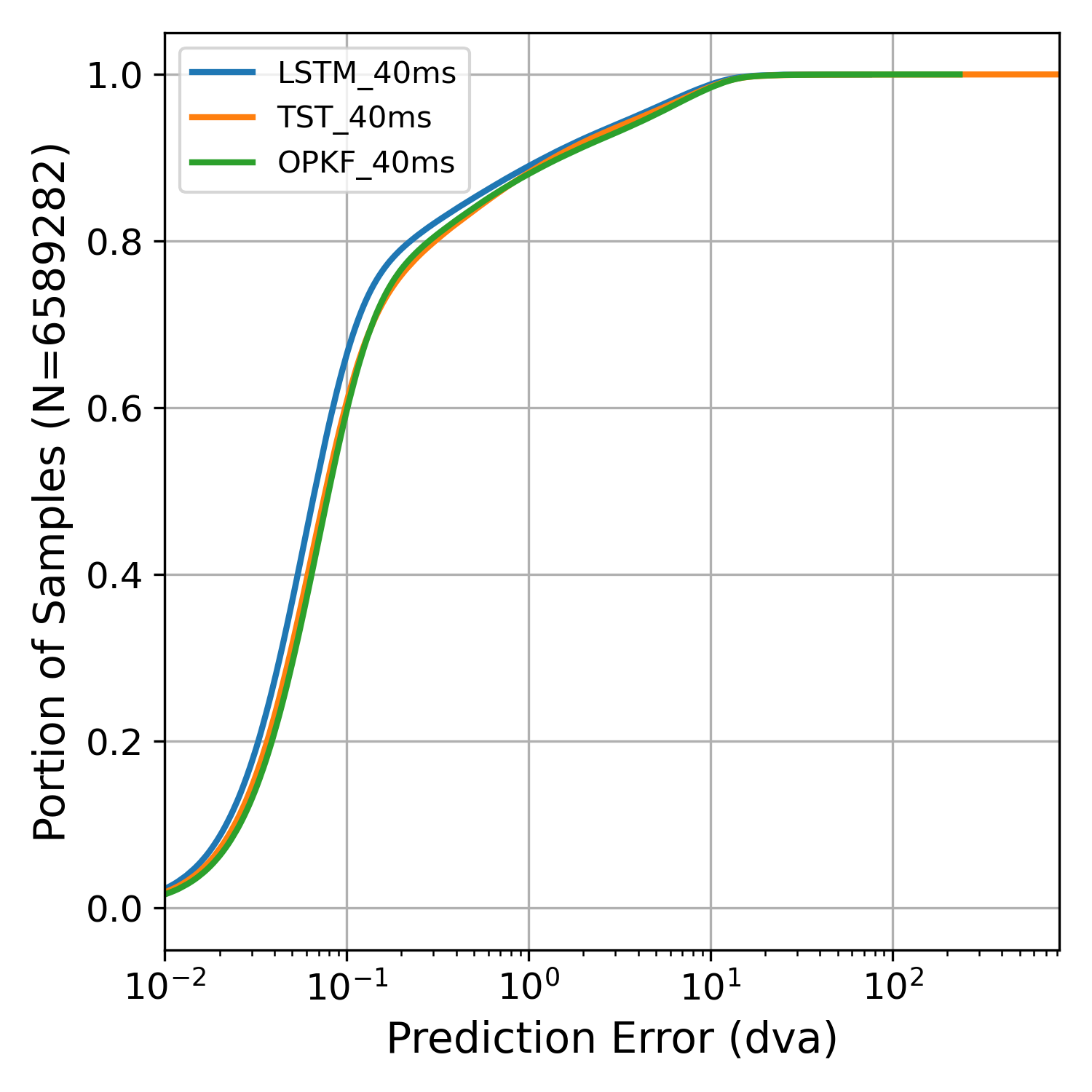}
        \caption{CDF of Event-Agnostic Errors}
        \label{fig:40_fullcdf}
    \end{subfigure}
    \hspace{0.02\textwidth} 
    \begin{subfigure}{0.31\textwidth}
        \includegraphics[width=\textwidth, height=4cm]{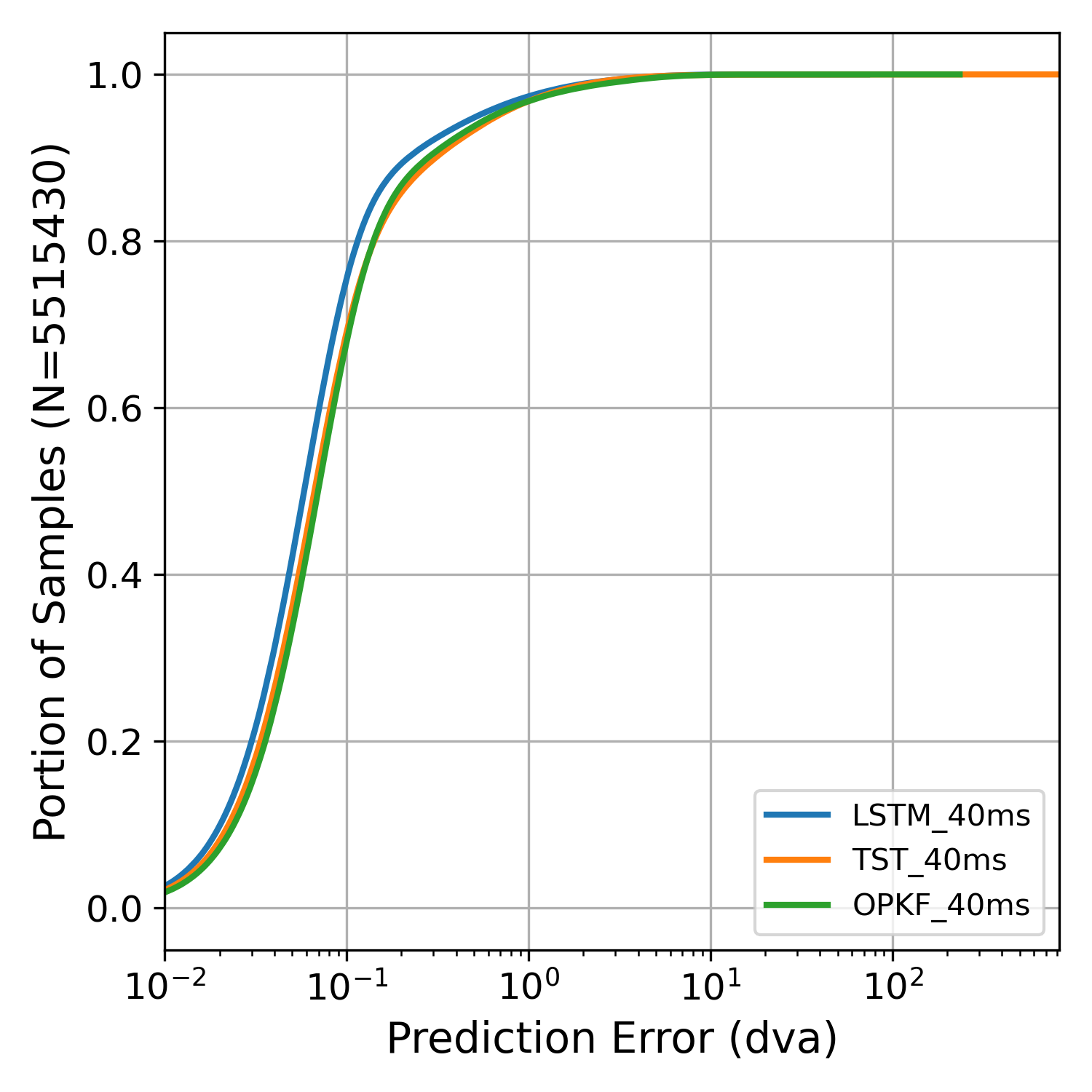}
        \caption{CDF of Fixation Errors}
        \label{fig:40_fullcdffix}
    \end{subfigure}
    \hspace{0.02\textwidth} 
    \begin{subfigure}{0.31\textwidth}
        \includegraphics[width=\textwidth, height=4cm]{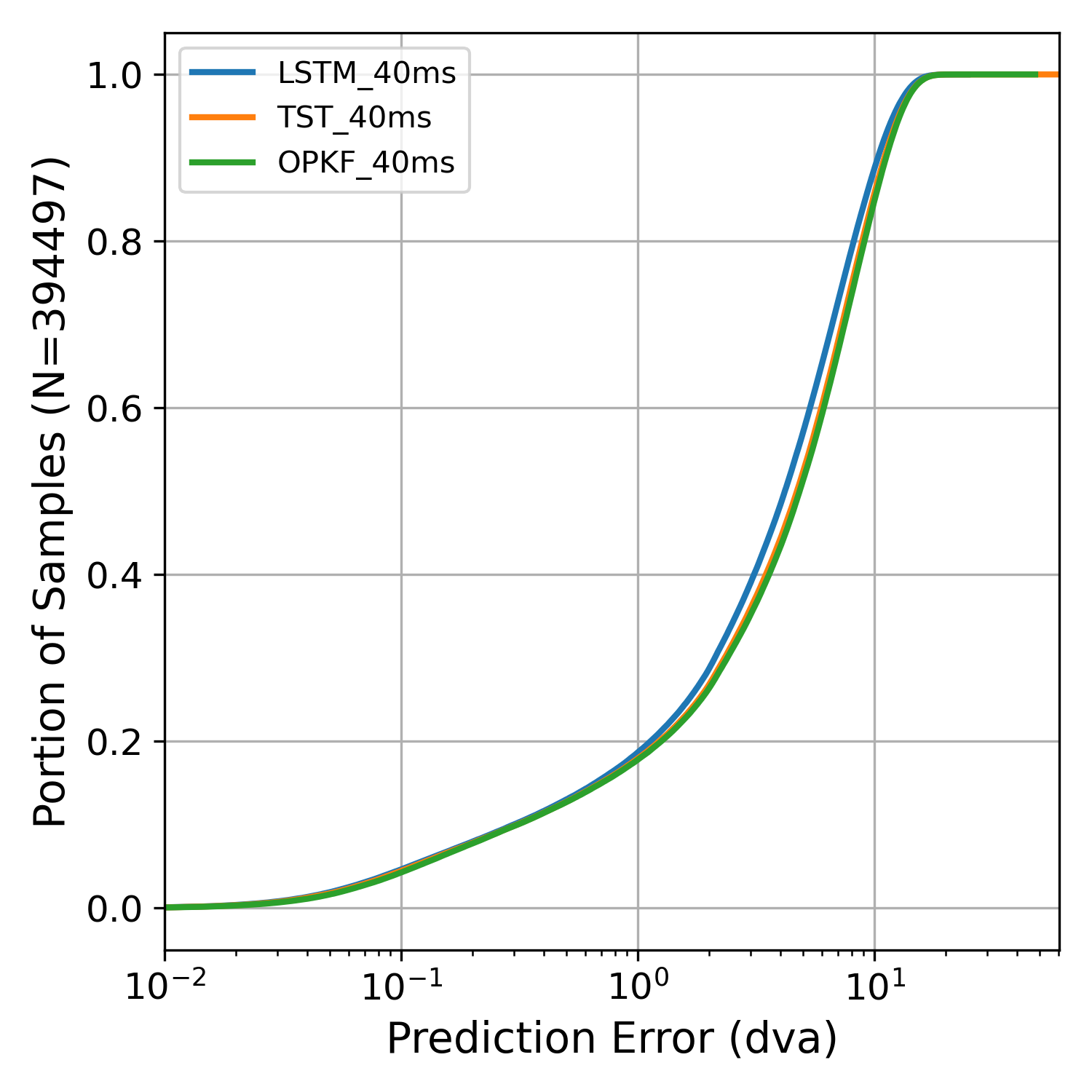}
        \caption{CDF of Large Saccade Errors}
        \label{fig:40_fullcdfsaclar}
    \end{subfigure}
    \vspace{0.01\textwidth} 
    \begin{subfigure}{0.31\textwidth} 
        \includegraphics[width=\textwidth, height=4cm]{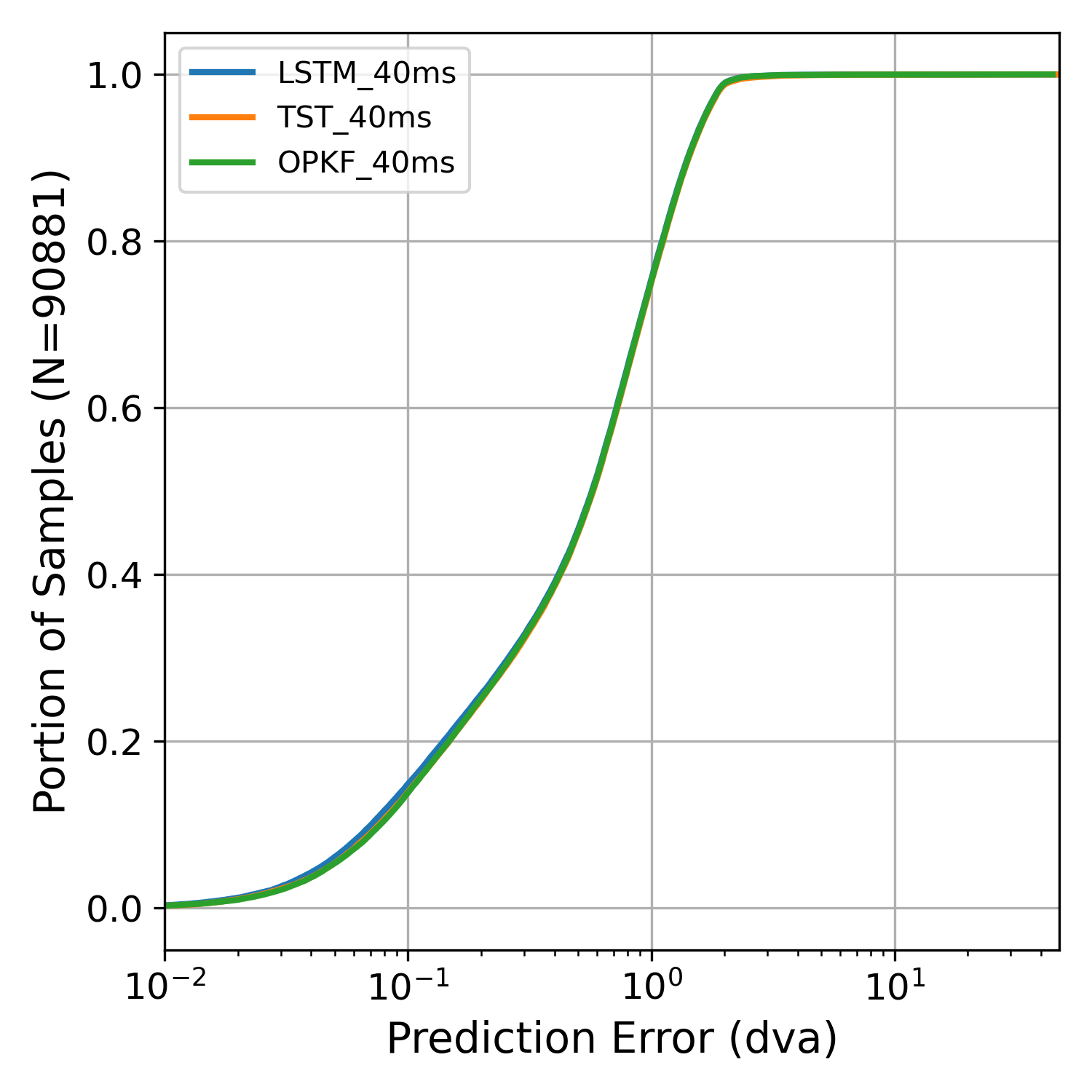}
        \caption{CDF of Small Saccade Errors}
        \label{fig:40_fullcdfsacsmall}
    \end{subfigure}
    \hspace{0.02\textwidth} 
    \begin{subfigure}{0.31\textwidth}
        \includegraphics[width=\textwidth, height=4cm]{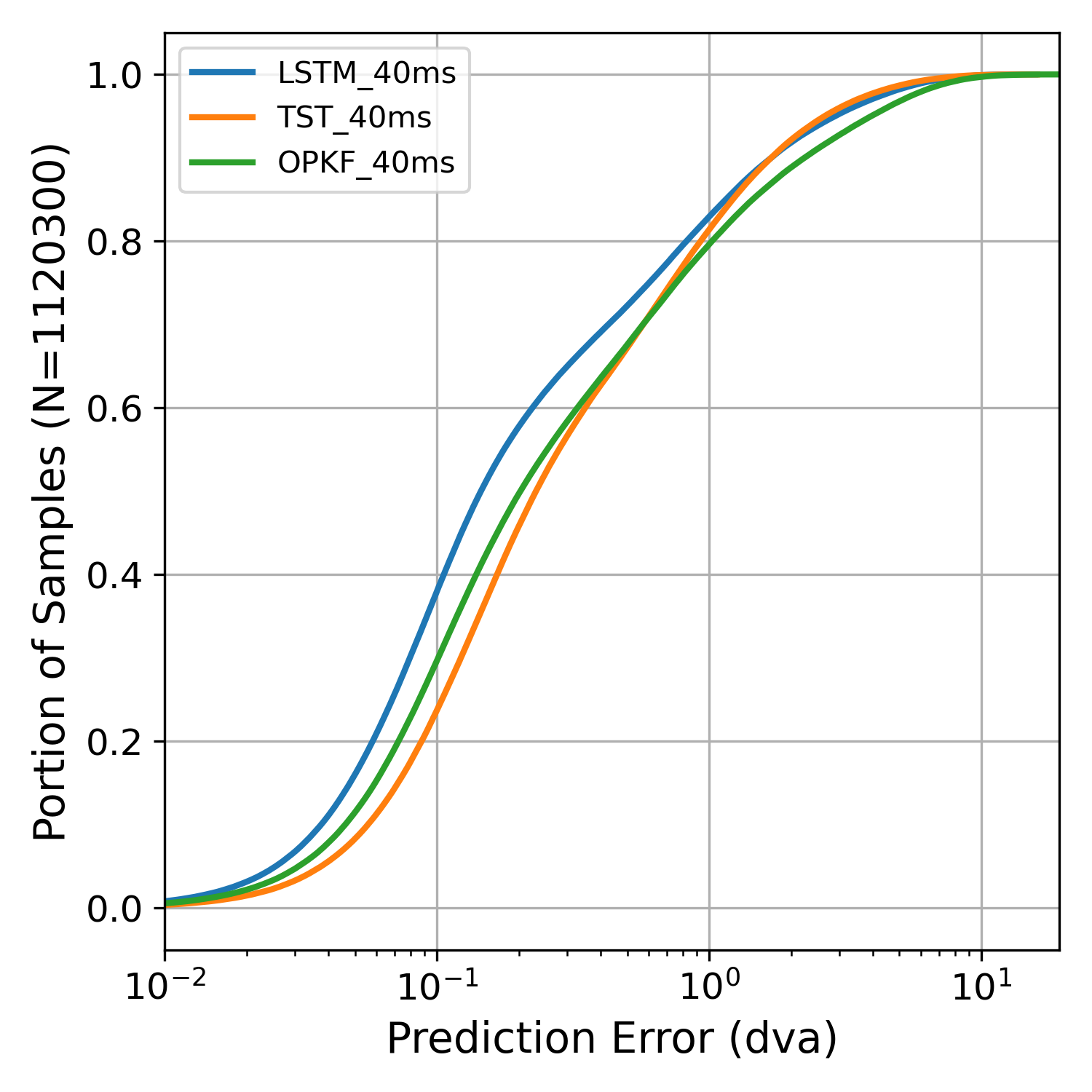}
        \caption{CDF of CEP Errors}
        \label{fig:40_fullcdfcep}
    \end{subfigure}
    \hspace{0.02\textwidth}
    \begin{subfigure}{0.31\textwidth}
        \includegraphics[width=\textwidth, height=4cm]{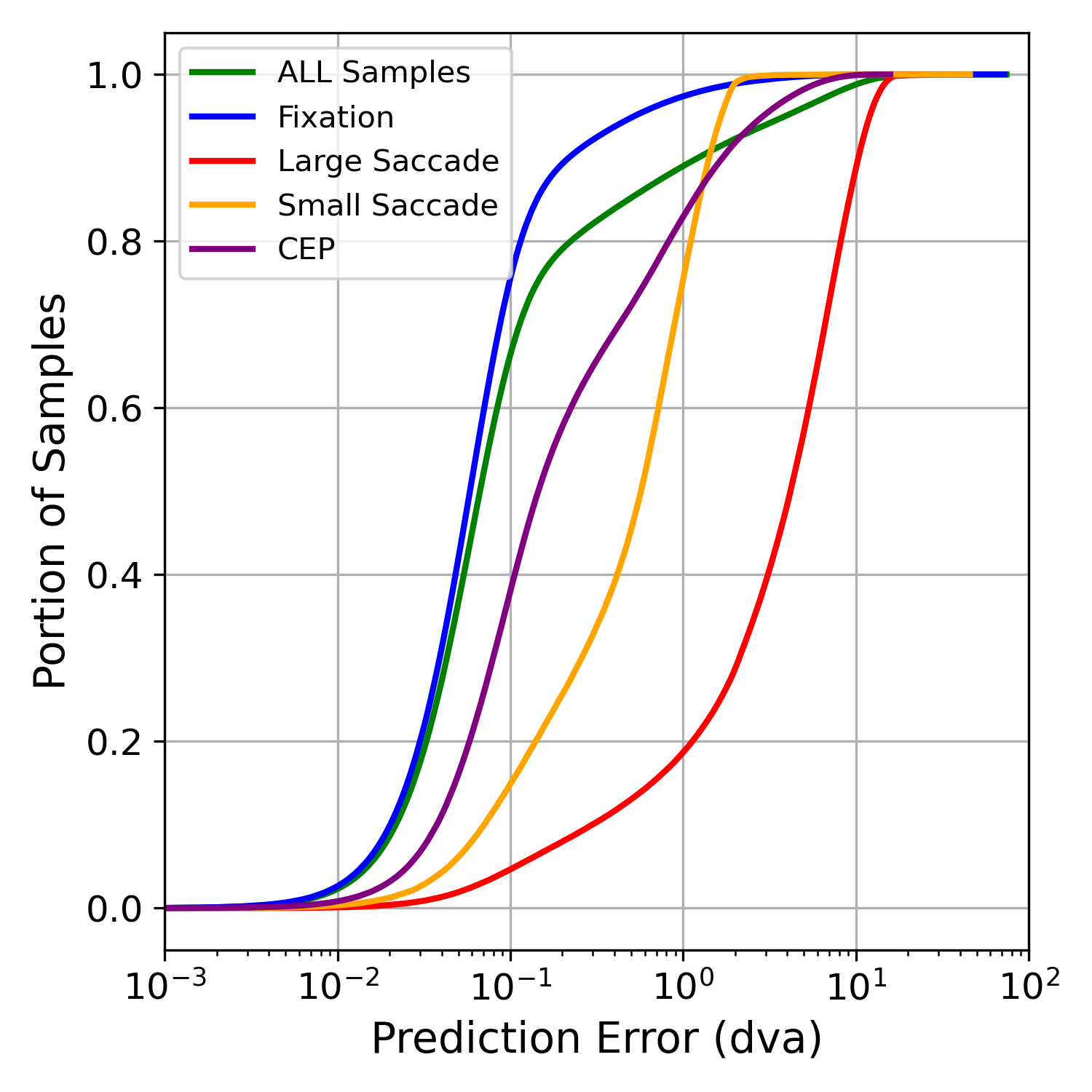}
        \caption{CDF of LSTM Errors of All Events}
        \label{fig:40_fullcdf_all}
    \end{subfigure}
    
    \label{fig:per_sample_pl}
    \vspace{-7mm}
\end{figure*}

\subsection{Gaze Prediction as Function of Eye-Movement Type}
In Figure \ref{fig:40_fullcdf}, we present the CDF across all the samples regardless of the eye movement classification. From this type of analysis, the model results are indistinguishable. Figures \ref{fig:40_fullcdffix}, \ref{fig:40_fullcdfsaclar}, \ref{fig:40_fullcdfsacsmall}, \ref{fig:40_fullcdfcep} are CDFs for fixations, large saccades, small saccades and CEP intervals. The curves for individual models look very similar for all eye movement types. In \ref{fig:40_fullcdf_all}, we plot just the LSTM model for all eye movement types. Gaze prediction performance is best within fixation, second best is all eye-movement-type, and the CEP interval is third best, followed by small and then large saccades.

\begin{figure}
    \centering
    \caption{Gaze Prediction as Function of Eye-Movement Type}
    \begin{subfigure}{0.32\textwidth}
        \includegraphics[width=\textwidth, height=4cm]{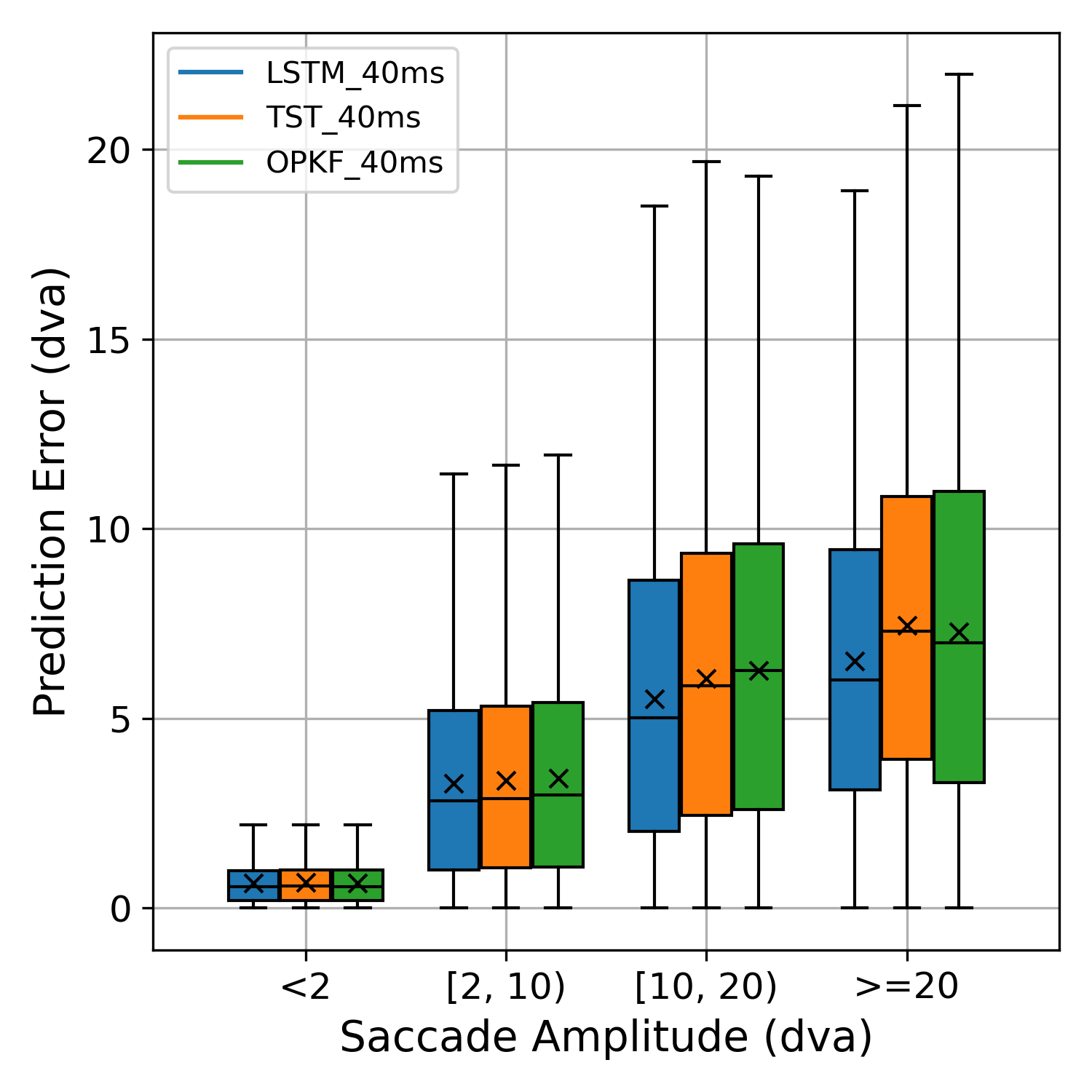}
        \caption{\centering Prediction Error As a Function of Saccade Size}
        \label{fig:40_sac_amp}
    \end{subfigure}
    \hfill
    \begin{subfigure}{0.32\textwidth}
        \includegraphics[width=\textwidth, height=4cm]{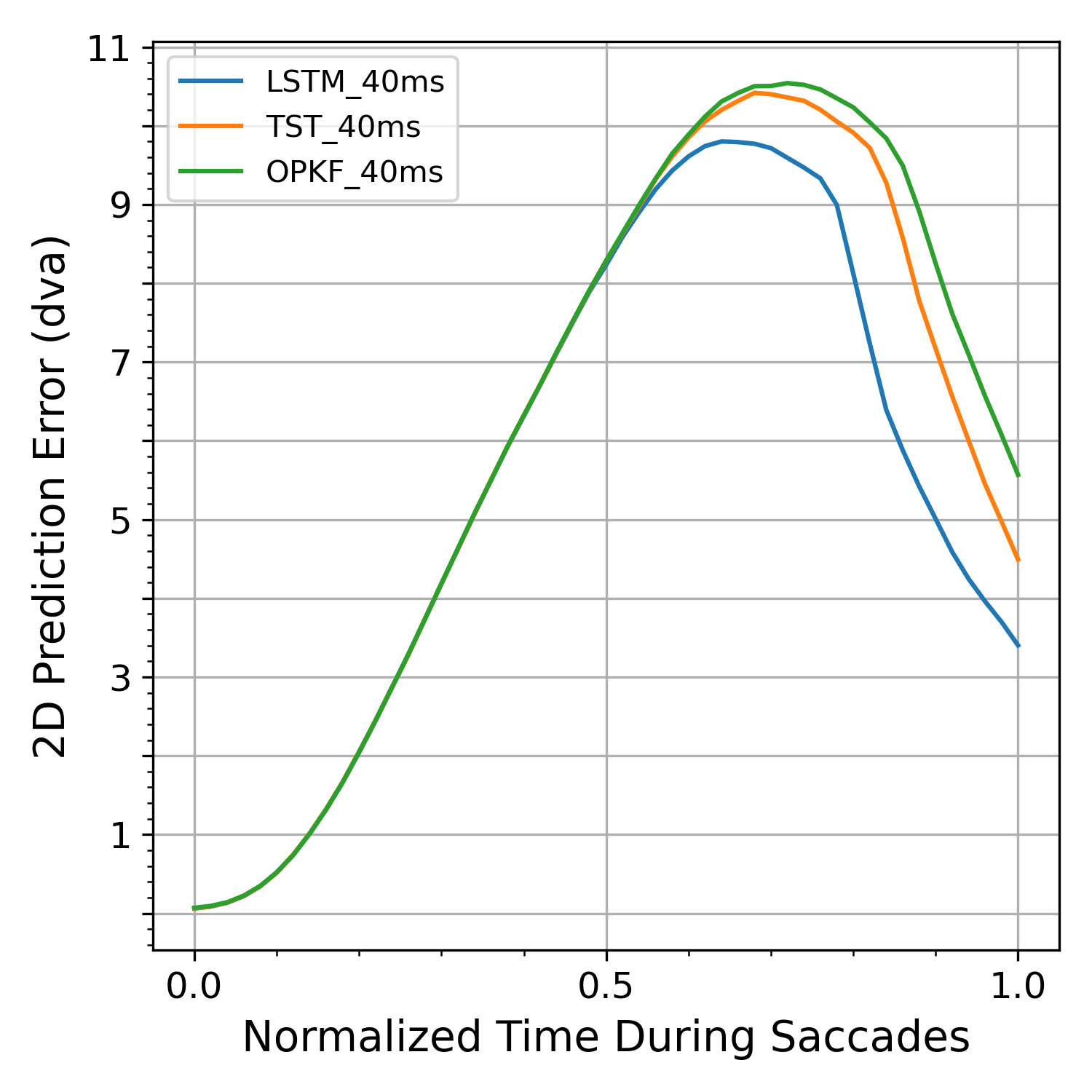}
        \caption{\centering Prediction Error as Function of Normalized Saccade Time}
        \label{fig:40_sac_10_20}
    \end{subfigure}
    \hfill
    \begin{subfigure}{0.32\textwidth}
        \includegraphics[width=\textwidth, height=4cm]{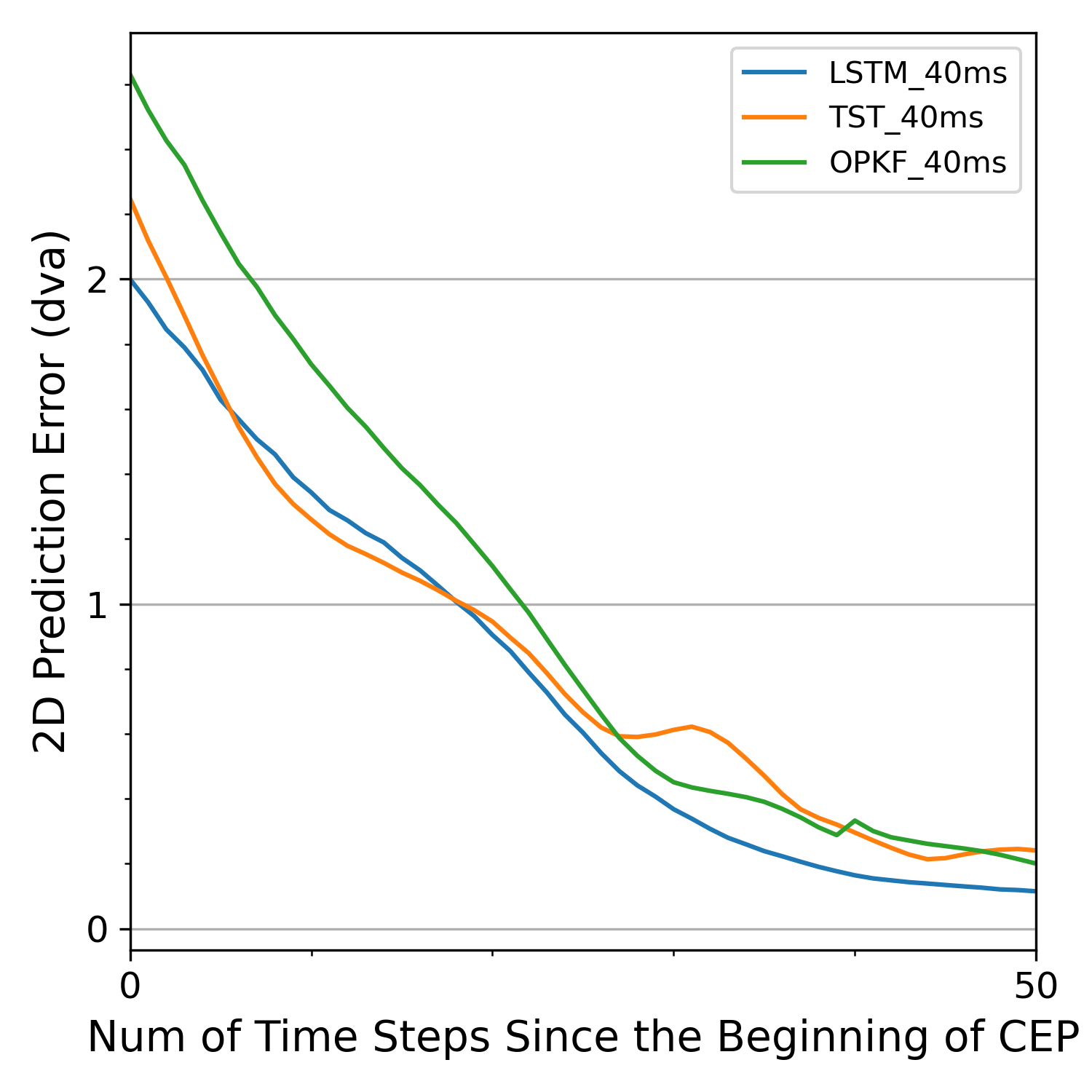}
        \caption{\centering Prediction Error During the First Half of CEP}
        \label{fig:40_cep_50}
    \end{subfigure}
    
    \label{fig:per_event_pl}
    \vspace{-3mm}
\end{figure}

We also analyzed the gaze prediction error as a function of the size of saccades (Fig. \ref{fig:40_sac_amp}). In Fig. \ref{fig:40_sac_10_20}, we present the prediction error as a function of normalized saccade time for saccades of amplitude $[10, 20]$. The most significant prediction error occurred in the middle of the saccade but began decreasing toward the end. In Fig. \ref{fig:40_cep_50}, the gaze prediction error also changes, i.e., declines, over time as the CEP samples approach the end of the first half of the CEP period.

\subsection{Individual Differences in Gaze Prediction Performance}

\begin{figure}
    \vspace{-4mm}
    \caption{Subject Profiles in Gaze Prediction Error}
    \begin{subfigure}{0.32\textwidth}
        \includegraphics[width=\textwidth]{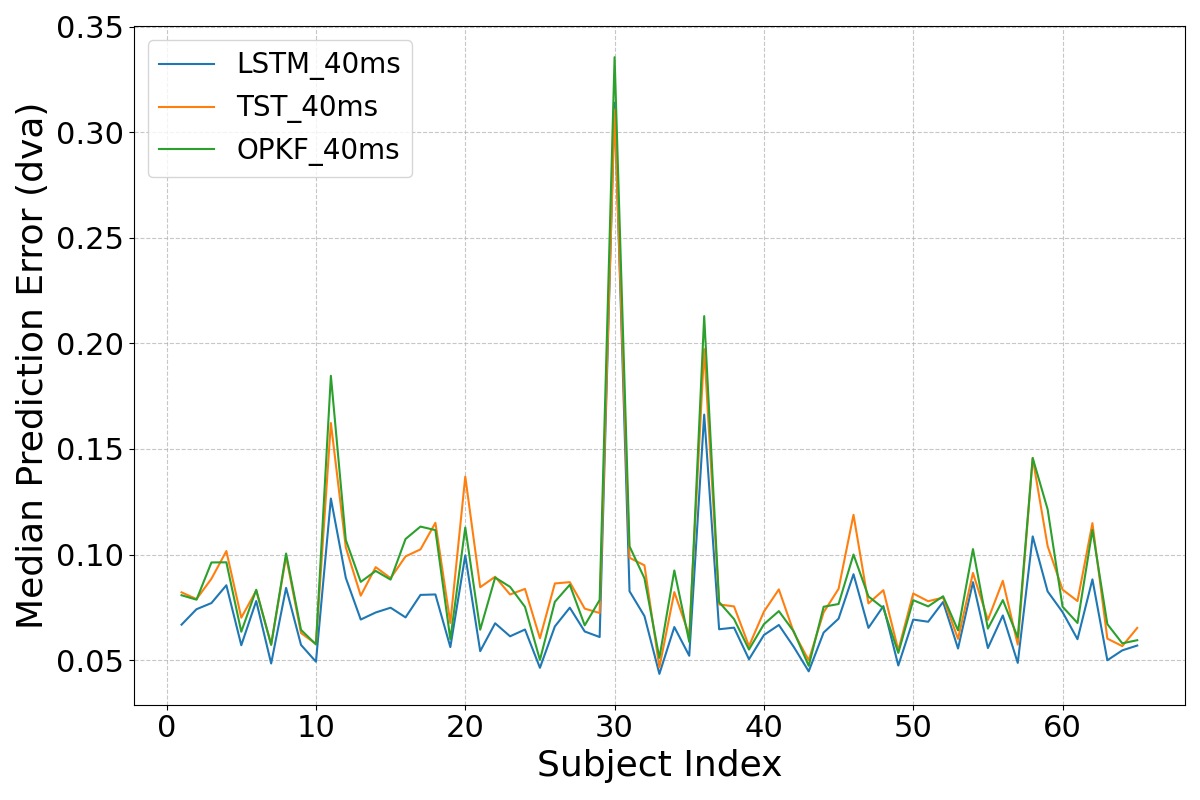}
        \caption{Fixation}
        \label{fig:diff_p50_fix}
        \vspace{-3mm}
    \end{subfigure}
    \hfill
    \begin{subfigure}{0.32\textwidth}
        \includegraphics[width=\textwidth]{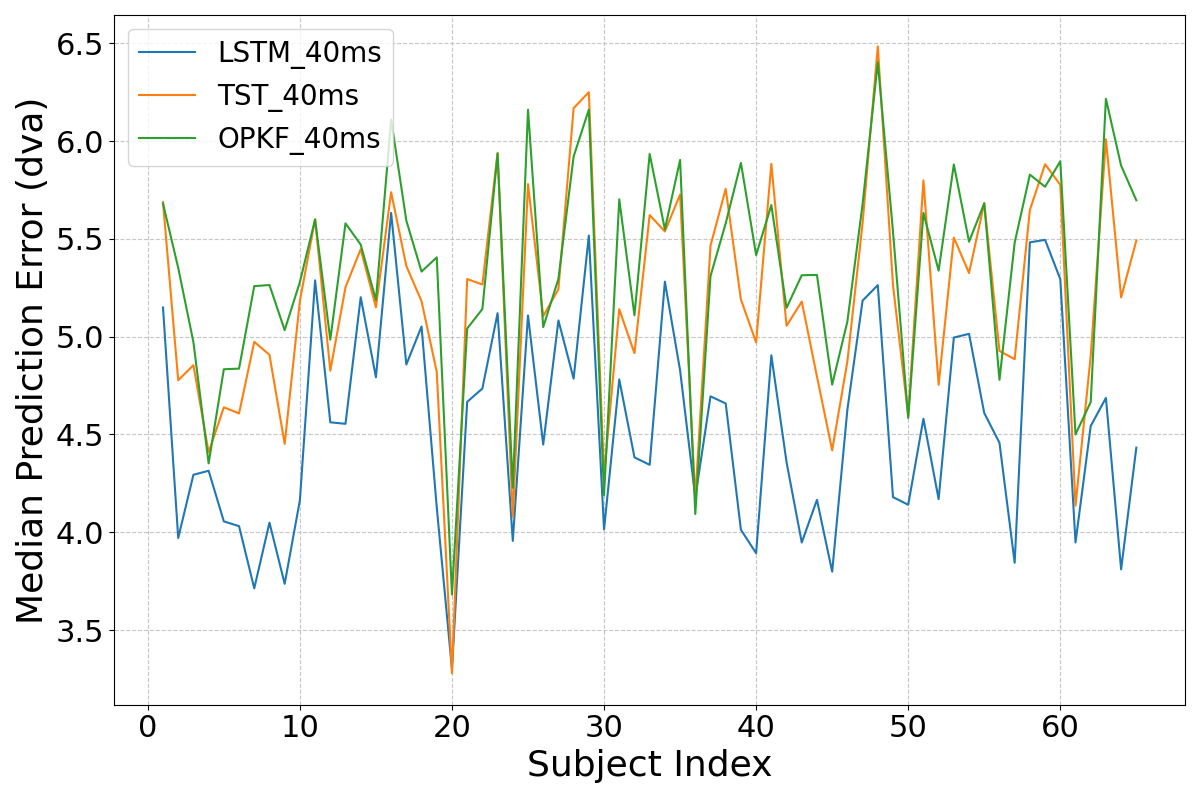}
        \caption{Large Saccades}
        \label{fig:diff_p50_lr_sac}
        \vspace{-3mm}
    \end{subfigure}
    \hfill
    \begin{subfigure}{0.32\textwidth}
        \includegraphics[width=\textwidth]{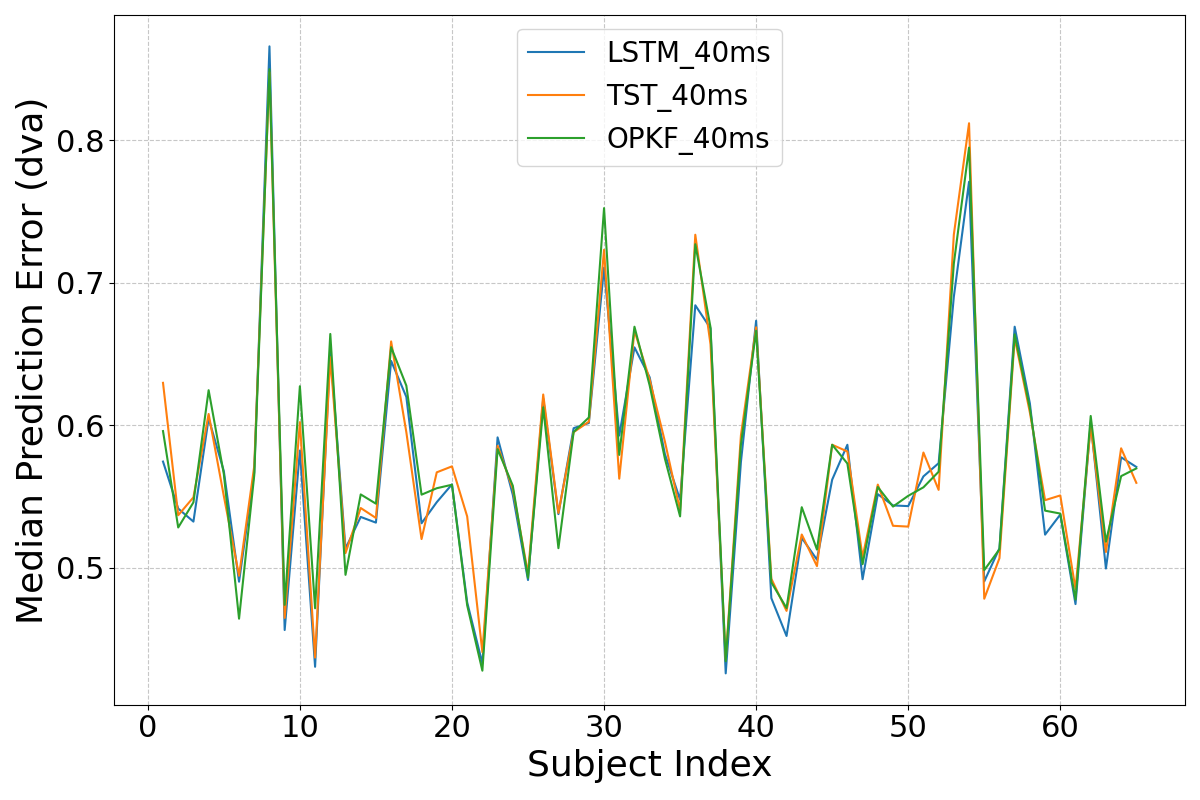}
        \caption{Small Saccades}
        \label{fig:diff_p50_sm_sac}
        \vspace{-3mm}
    \end{subfigure}
    \label{fig:diff_sub_plots}
    \vspace{-2mm}
\end{figure}

We start by creating a profile of median error across subjects (Fig.\ref{fig:diff_sub_plots}) for fixations (Fig.\ref{fig:diff_p50_fix}), large saccades (Fig.\ref{fig:diff_p50_lr_sac}), and small saccades (Fig.\ref{fig:diff_p50_sm_sac}). For fixations and small saccades, the profiles across subjects appear quite similar across models. For large saccades, the individual differences in profiles are strongly influenced by the model applied. All three algorithms use fundamentally different approaches to learning from training data. However, at least for fixations and small saccades, they achieve roughly the same median prediction per participant. These results suggest correlating individual oculomotor measures with gaze prediction performance (see Section \ref{sec:corr_analysis} below).

Table \ref{tab:stat_combined} presents some statistics on individual differences during fixation, large and small saccades. In this table, ``Ratio'' is the ratio of the maximum value to the minimum value. The LSTM model's maximum value for fixations is over 7 times larger than the minimum value. The other gaze prediction models have similar ratios for fixations. The ratios for large saccades are all less than 2, meaning that the maximum is $\approx$ double the size of the minimum value. This is also true for the small saccades. The interquartile ranges (``IQR'') provide another measure of variability across subjects in gaze prediction performance. In the case of fixation, the IQRs are the smallest during fixations. They are also very small for the saccades of smaller amplitudes but quite large for large saccades.

\subsection{What Oculomotor Indices Predict Individual Differences in Gaze Performance}
\label{sec:corr_analysis}

We performed a correlation analysis to determine what specific oculomotor features predict individual differences. We start with the set of extremely detailed and very specific oculomotor features presented in \cite{rigas2018study}. There are over 1000 such features, which would create a massive multiple-comparison problem. So, we chose only the subset of these features that concern fixation and saccades, which refer to a radial measure (as opposed to horizontal or vertical position) and represent a median across eye movement all fixation or all saccades. It left us with 35 oculomotor features. In addition, we wanted to correlate the fixation noise threshold as a general measure of velocity noise during fixation \cite{mnh}. To compute this, we create a histogram of radial velocity for every sample, which is part of a fixation. The fixation velocity threshold is the \nth{90} percentile of this distribution for each subject. We also correlate 2 data quality measures (accuracy and precision \cite{lohr2019eval}). No significant correlation scores were noted for small saccades, so these analyses are not presented.

\begin{table}
    \centering
    \begin{tabular}{|l|c|c|c|c|c|c|c|c|c|c|c|c|} \hline
        \multirow{2}{*}{Model} & \multicolumn{4}{c|}{Fixation}        & \multicolumn{4}{c|}{Large Saccades}  & \multicolumn{4}{c|}{Small Saccades} \\ \cline{2-13}
            & Min   & Max   & Ratio & IQR         & Min   & Max   & Ratio & IQR         & Min   & Max   & Ratio & IQR         \\ \hline
        LSTM                   & 0.044 & 0.314 & 7.21  & 0.021       & 3.3   & 5.632 & 1.71  & 0.868       & 0.426 & 0.866 & 2.03  & 0.081       \\ \hline
        TST                    & 0.047 & 0.311 & 6.67  & 0.025       & 3.279 & 6.484 & 1.98  & 0.767       & 0.436 & 0.848 & 1.94  & 0.079       \\ \hline
        OPKF                   & 0.047 & 0.335 & 7.09  & 0.032       & 3.682 & 6.403 & 1.74  & 0.648       & 0.428 & 0.850 & 1.99  & 0.093       \\ \hline
    \end{tabular}
    \caption{Error Distribution Statistics Per Fixation and Saccade Type}
    \label{tab:stat_combined}
    \vspace{-5mm}
\end{table}

Since the data were generally not normally distributed, we report the Spearman correlation coefficient ($r_{s}$). Note that all of the $r_{s}$ in this section are still statistically significant after Bonferroni correction for multiple comparisons. As seen in Table \ref{tab:corr_combined}, The Fixation Noise Threshold significantly correlates with the median performance of all three models. Therefore, the noisier the fixation periods, the poorer the gaze prediction performance. In Table \ref{tab:corr_combined}, two measures were correlated with the median performance of models based on only large saccades. These oculomotor measures are defined below.
\vspace{-2mm}

\begin{table}
    \centering
    \begin{tabular}{|l||c|c|c||c|c|c|} \hline
        \multirow{2}{*}{Model} & \multicolumn{3}{|c||}{Fixation Analysis} & \multicolumn{3}{c|}{Saccade Analysis} \\ \cline{2-7}
        & Feature                & $r_{s}$ & $p$-value  &          Feature                & $r_{s}$ & $p$-value  \\ \hline
        LSTM & FixNoiseThr           & 0.79    & $p < 10^{-4}$       & PkVelDurRatioRMd & 0.75    & $p < 10^{-4}$  \\ \hline
        TST & FixNoiseThr           & 0.85    & $p < 10^{-4}$       & MnVelRMd           & 0.65    & $p < 10^{-4}$  \\ \hline
        OPKF  & FixNoiseThr           & 0.93    & $p < 10^{-4}$       & PkVelDurRatioRMd & 0.59    & $p < 10^{-4}$  \\ \hline
    \end{tabular}
    \caption{Correlation of Oculomotor Measures and Median Prediction Errors}
    \label{tab:corr_combined}
    \vspace{-5mm}
\end{table}

\subsubsection{PkVelDurRatioRMd}
First, we compute the peak radial velocity for every saccade. It is then divided by the number of samples in each saccade. So, this becomes a measure of peak velocity per millisecond for each saccade. $PkVelDurRatioRMd$ is the median value of this measure across saccades. It is a measure of velocity in saccades. The data in Table \ref{tab:corr_combined} indicate that gaze prediction performance decreases as the velocity during saccades increases. 
\vspace{-2mm}

\subsubsection{MnVelRMd}
This measure is simply the mean of the median velocity during saccades. It correlates highly with the previous measure ($PkVelDurRatioRMd$: $r_s$ = 0.86, ($p-value$ is effectively zero)). These two measures are nearly identical. The data in Table \ref{tab:corr_combined} indicate that gaze prediction performance decreases as the mean velocity during saccades increases. Similar correlations were noted for other prediction intervals.  

\section{Discussion}
The main findings of this study are: (1) that gaze prediction performance is different for different eye movement types; (2) that there are important individual differences in gaze prediction performance; and (3) that several oculomotor measures correlate with reduced gaze prediction performance.  

Regarding eye movement types, the order of lowest to highest error was as follows: fixation, CEP intervals, small saccades, and then large saccades. Considering that the eye is mostly stable during fixation, it follows that gaze prediction performance was high. During the CEP period, the eye is also more stable than during saccades, so it follows that this interval had comparatively good gaze prediction. During saccades, the eye is rapidly moving, and it follows that gaze prediction performance is poor. Also, we found that gaze prediction performance is a function of saccade amplitude, with larger saccades having a more significant error. The gaze prediction error is largest during the initial portion of the CEP period and reduces markedly toward the end of the CEP period \cite{aziz2023practical}. 

The individual differences in median error profiles were apparent. The range analysis, the ratio analysis, and IQR statistics all document subject-to-subject variation. The range analysis is sensitive to outliers, whereas the IQR analysis is not. The IQRs were much larger for large saccades than for small saccades and fixation. Thus, there are more considerable individual differences in the gaze prediction of saccades.

For fixations, we found that a measure of fixation velocity noise was highly correlated with gaze prediction performance. For saccades, we found that a measure of saccade velocity was correlated with gaze prediction performance. We propose that future studies might include such oculomotor features to potentially reduce individual differences.

We propose that future studies of gaze prediction should include analyses of individual differences. This is because prediction algorithms need to work well for all subjects and mean performance obscures important variation. Some effort to tune models to reduce individual differences might be an important strategy going forward.

We found that our estimate of fixation noise (Fixation Noise Threshold \cite{mnh}) was highly correlated with poorer gaze prediction performance during fixation. This makes perfect sense. Obviously, the more fixation noise, the harder it is to predict future gaze position. We also found that subjects with higher velocity saccades also had poorer gaze prediction performance. This also makes perfect sense: the faster the saccade, the more difficult it was to predict the future eye position.

Previous research on the use of eye movements for biometric identification has been relatively successful and has produced a set of 128 embeddings \cite{lohr2022eye}. These embeddings contain information reflecting individual differences in oculomotor performance. One future goal might be to employ these embeddings into the gaze prediction pipeline. 

Our dataset is atypical for gaze prediction studies due to its greater accuracy and temporal precision (our sample rate was 1000 Hz, based on a state-of-the-art eye-tracking device). OPKF can work with any quality signal by its design, even if irregularities occur in real time. For deep learning models, higher signal noise will pose challenges during real-time gaze prediction. Our analyses presented herein are a reasonable template for future deep-learning approaches to lower-quality data.

All of this work employed only eye position data. Studies based on entirely different input signals (e.g., silency maps or head position) may have different results from ours. The present study should be relevant to other studies that use gaze positions and other inputs, but each study will be based on unique signal quality characteristics. Our plans include evaluating lower-quality eye-position data from VR headsets.

\section{Conclusion}
In conclusion, we highlight our findings regarding individual differences. It is not enough for a model to work well on average. If a participant's gaze cannot be predicted, foveated rendering may not succeed. Gaze prediction may also enhance the Human-Computer-Interface experience, but this improvement may disappear if prediction is poor for some subjects. We propose that information regarding individual differences should be included in future models. Also, oculomotor metrics regarding fixation noise and saccade speed can be input to reduce the range of individual differences. A measure of individual differences, such as the interquartile range of various prediction errors, should be more commonly reported.

\bibliographystyle{unsrt}


\setcounter{figure}{0}
\setcounter{section}{0}
\setcounter{table}{0}

\appendix
\label{sec:appendix}
\newpage
Appendix

\section{Evaluating Gaze Prediction Performance as a Function of Prediction Interval}
In Appendix Figure \ref{fig:cdf_all_LSTM_25_40_60}, we present the CDFs for all models for all PIs for all eye-movement types for the LSTM model only. It is clear in this figure that shorter PIs are associated with less error. The exact same results can be observed in Figure \ref{fig:cdf_fix_LSTM_25_40_60}.

\begin{figure*}[h!]
    \caption{CDF Plots Across All Eye-Movement Events and Fixations Only}
    \centering
    \begin{subfigure}{0.33\textwidth} 
        \includegraphics[width=\textwidth, height=4.1cm]{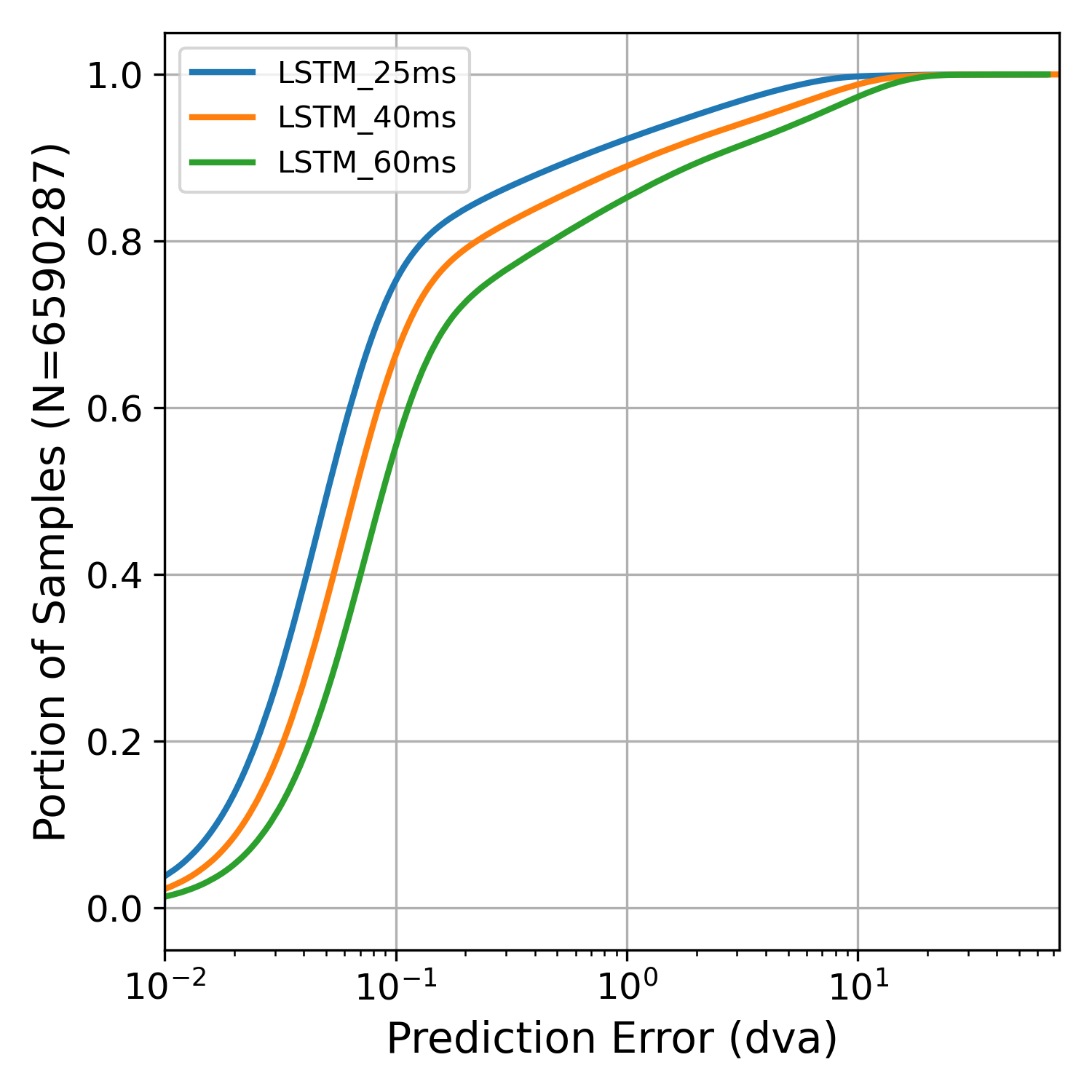}
        \caption{CDF of Event-Agnostic Errors LSTM}
        \label{fig:cdf_all_LSTM_25_40_60}
    \end{subfigure}
    \hspace{0.05\textwidth} 
    \begin{subfigure}{0.33\textwidth} 
        \includegraphics[width=\textwidth, height=4.1cm]{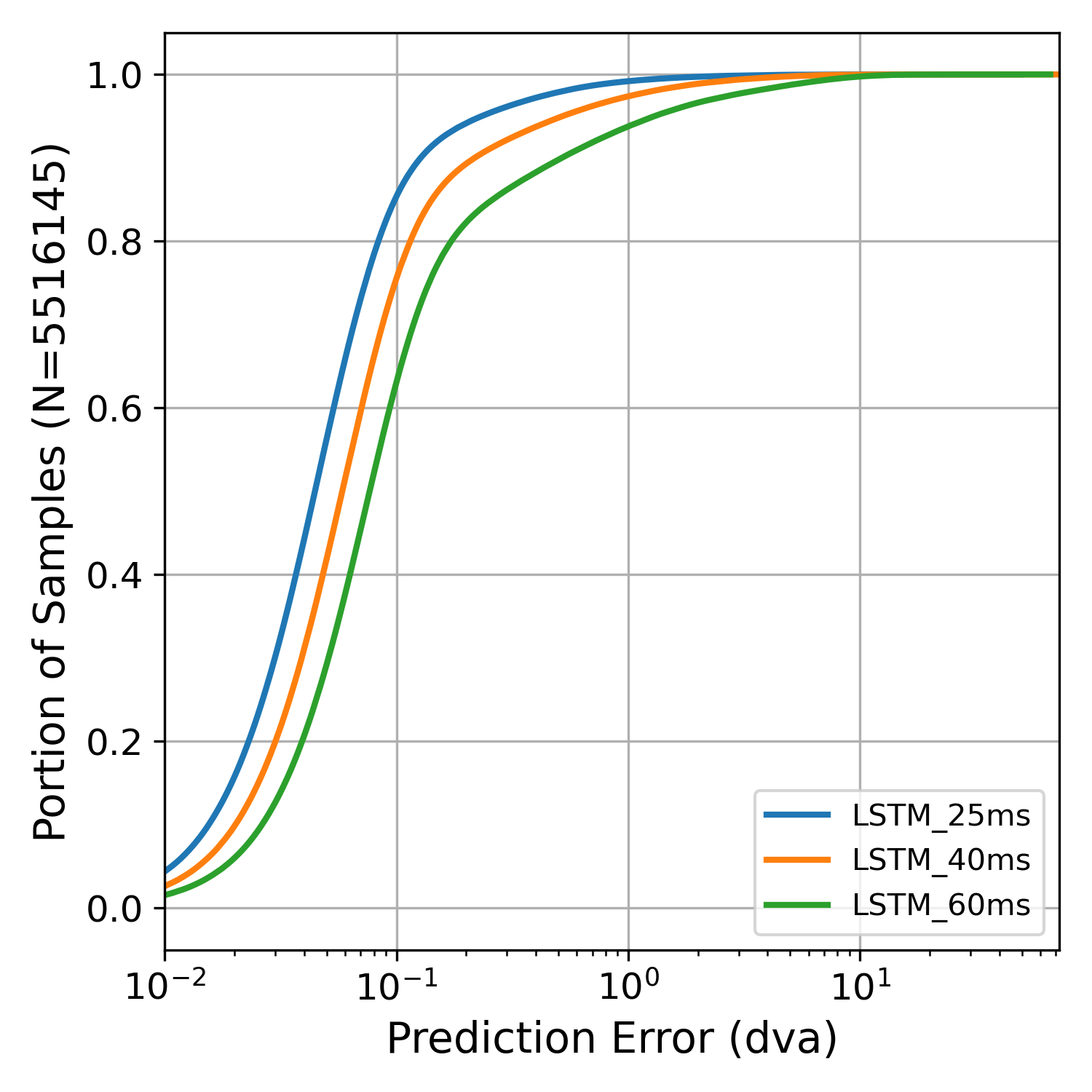}
        \caption{CDF of Fixation Errors LSTM}
        \label{fig:cdf_fix_LSTM_25_40_60}
    \end{subfigure}
    \label{fig:cdf_all_fix}
\end{figure*}

\section{Evaluating Gaze Prediction Performance as a Function of Eye Movement Type}
\begin{figure*}
    \begin{subfigure}{0.32\textwidth}
        \includegraphics[width=\textwidth, height=3.9cm]{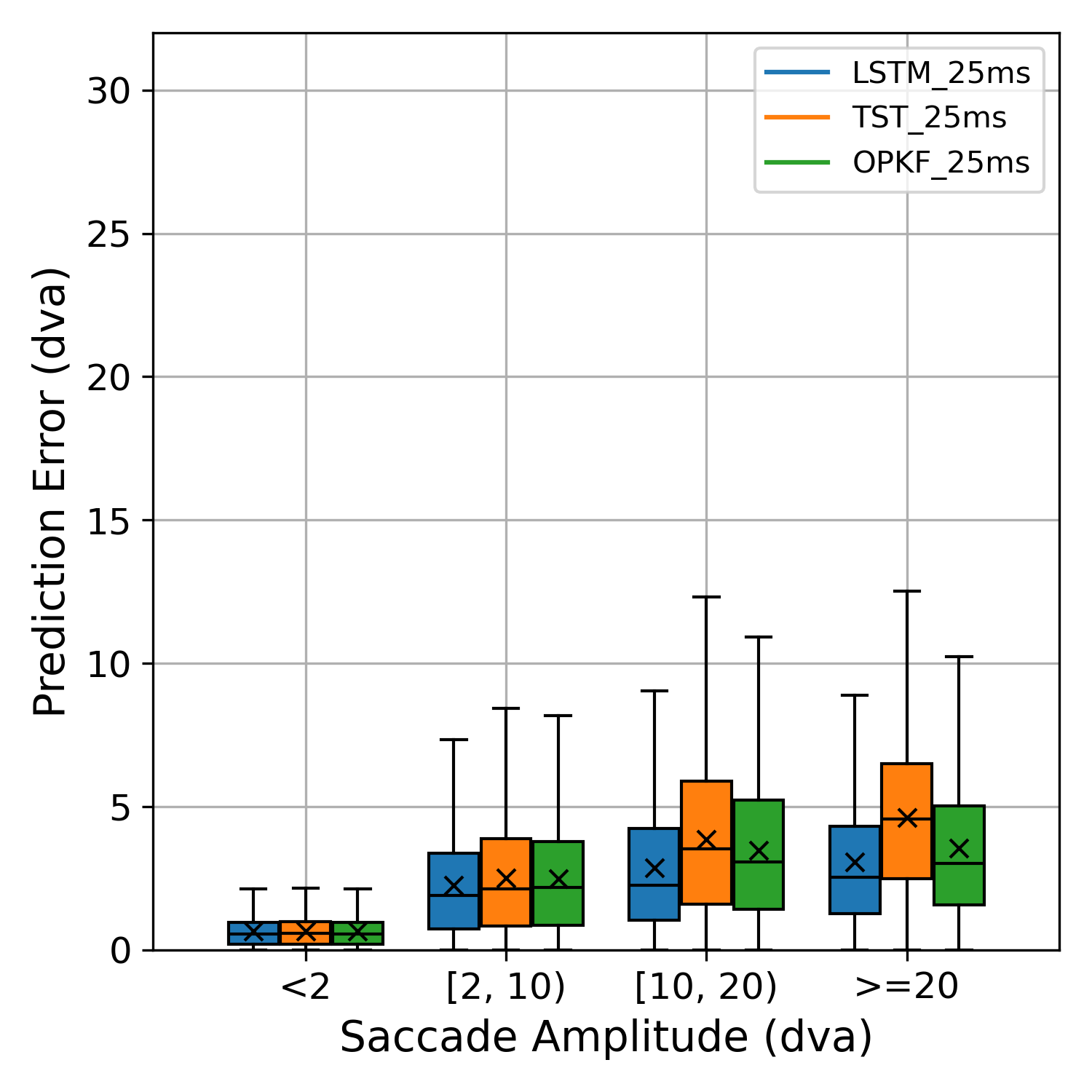}
        \caption{25 PI}
        \label{fig:app_sac_by_amp_25}
    \end{subfigure}
    \hfill
    \begin{subfigure}{0.32\textwidth}
        \includegraphics[width=\textwidth, height=3.9cm]{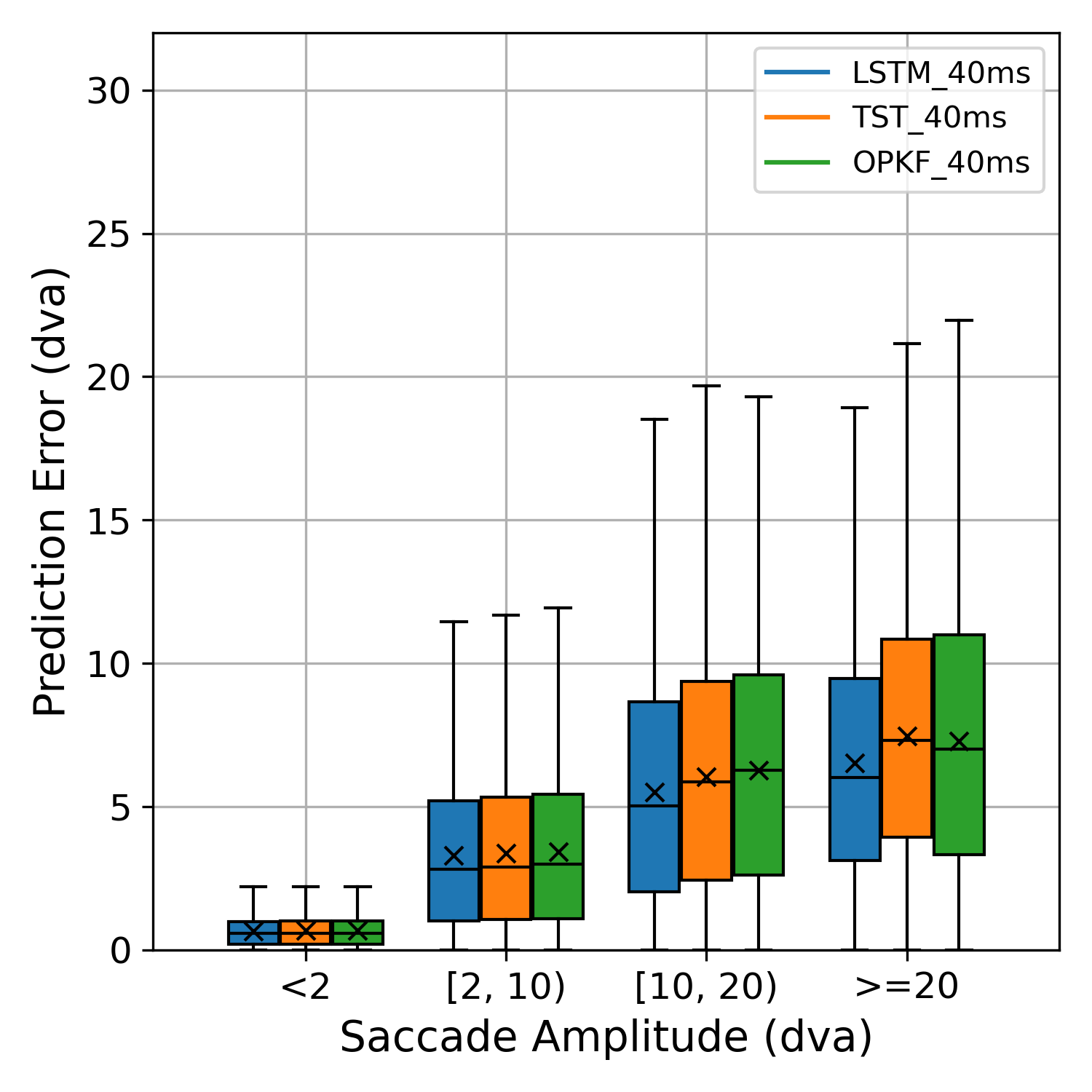}
        \caption{40 PI}
        \label{fig:app_sac_by_amp_40}
    \end{subfigure}
    \hfill
    \begin{subfigure}{0.32\textwidth}
        \includegraphics[width=\textwidth, height=3.9cm]{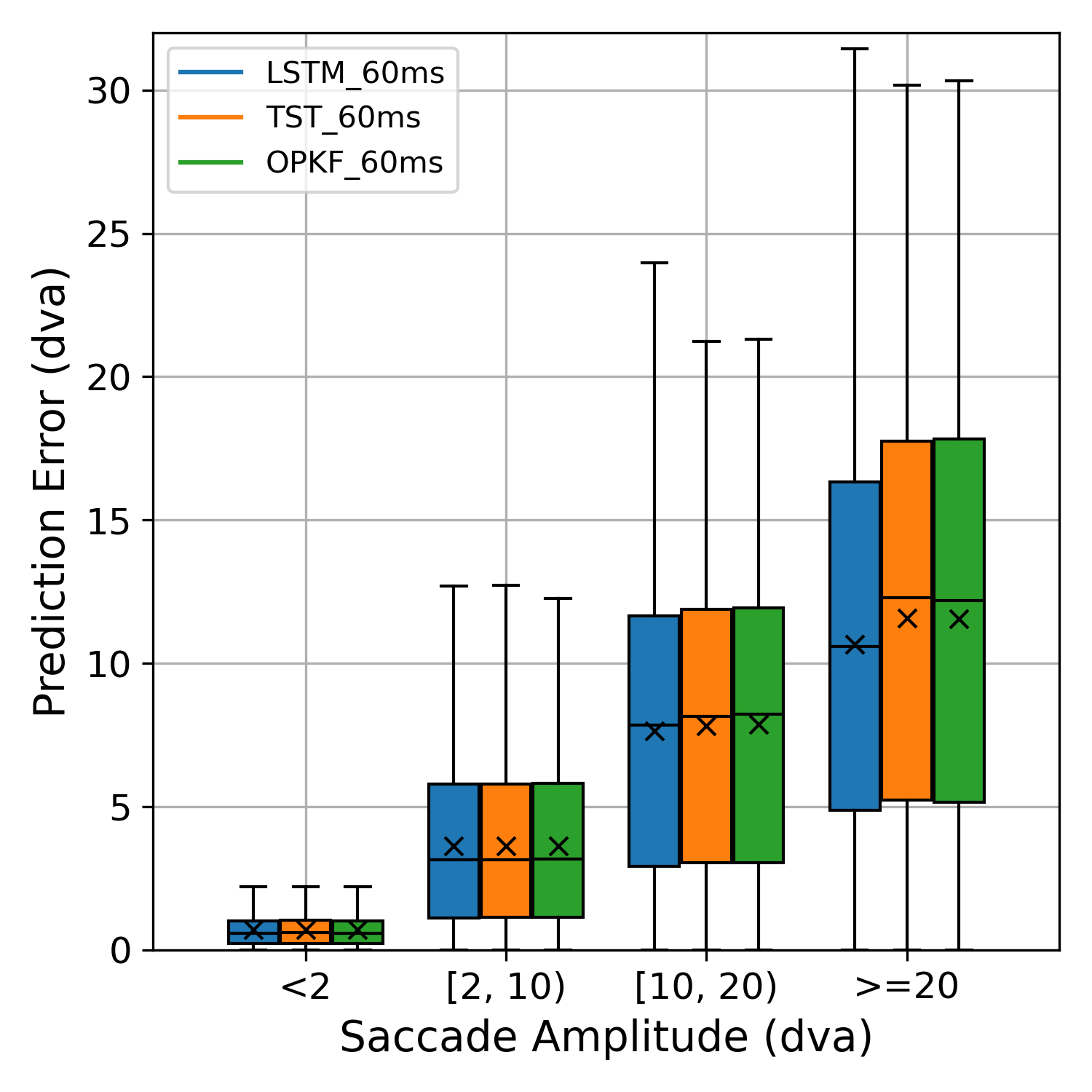}
        \caption{60 PI}
        \label{fig:app_sac_by_amp_60}
    \end{subfigure}
    \caption{Gaze Prediction Error as a Function of Saccade Size Across PIs}
    \label{fig:app_sac_by_amp}
\end{figure*}

In Figure \ref{fig:app_sac_by_amp}, we present boxplots representing prediction error as a function of saccade amplitude. Prediction error increases as a function of saccade amplitude. This effect is present for all three PIs. There is an increase in error for longer PIs.

\begin{figure*}
    \begin{subfigure}{0.32\textwidth}
        \includegraphics[width=\textwidth, height=3.9cm]{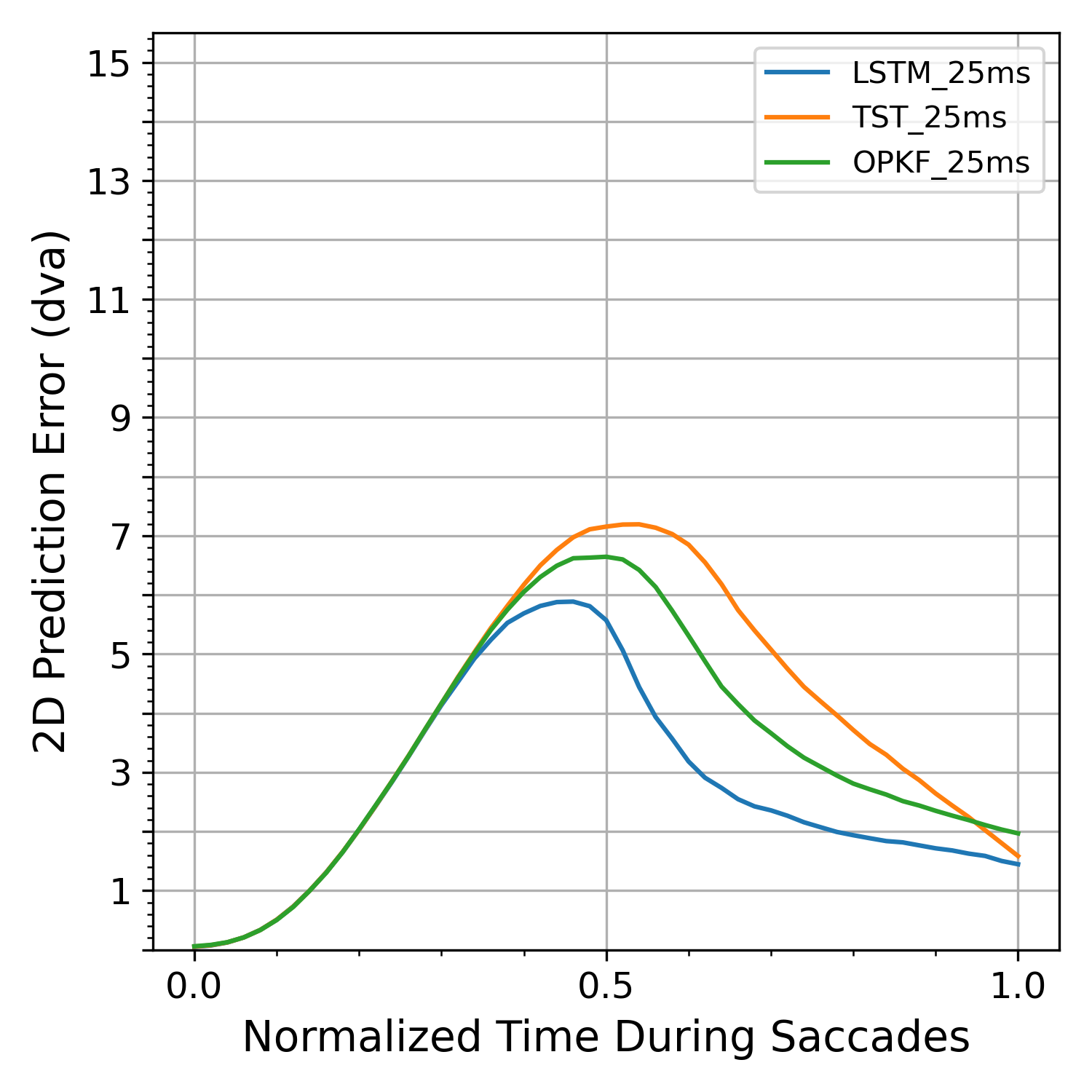}
        \caption{25 PI}
        \label{fig:app_25_sac_stat}
    \end{subfigure}
    \hfill
    \begin{subfigure}{0.32\textwidth}
        \includegraphics[width=\textwidth, height=3.9cm]{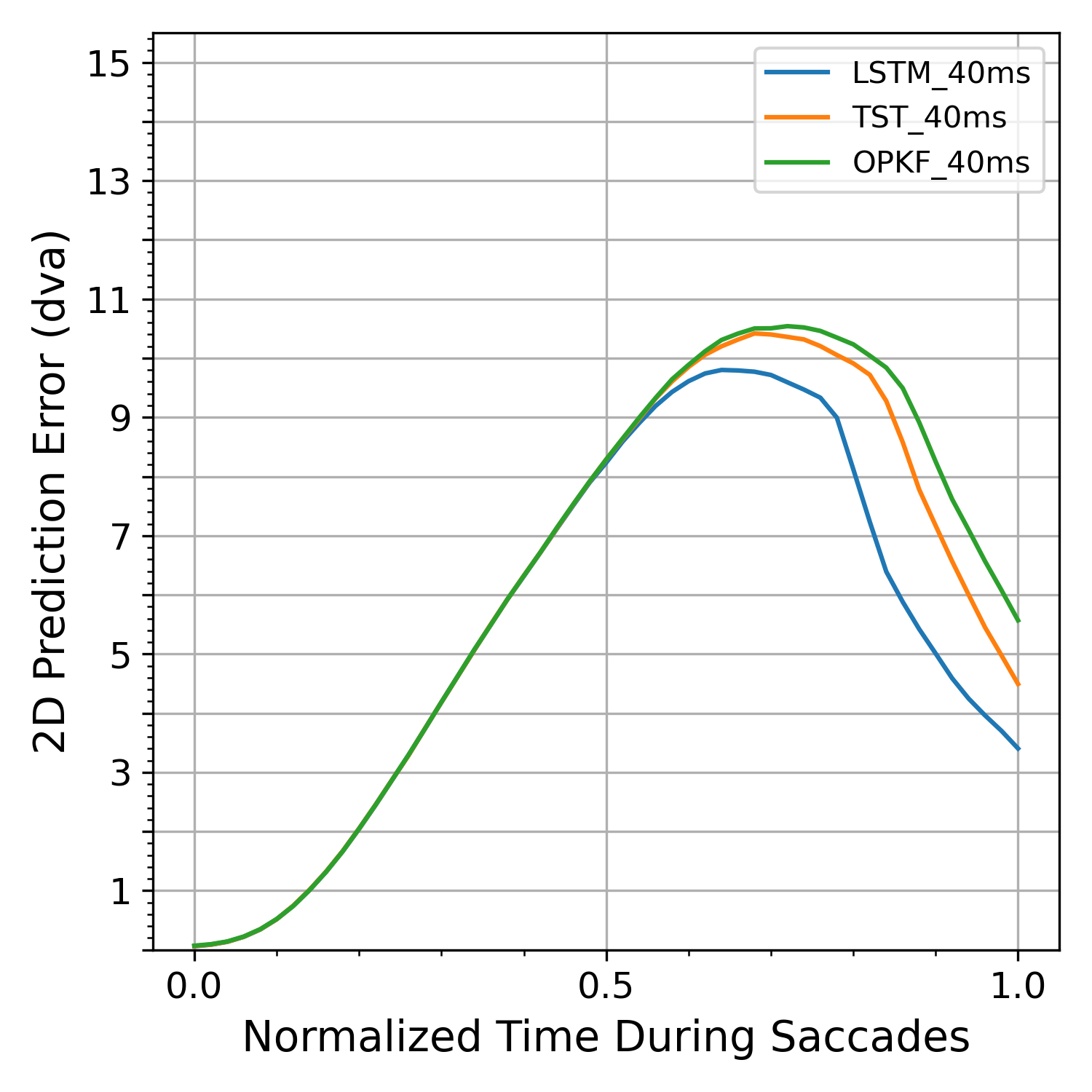}
        \caption{40 PI}
        \label{fig:app_40_sac_stat}
    \end{subfigure}
    \hfill
    \begin{subfigure}{0.32\textwidth}
        \includegraphics[width=\textwidth, height=3.9cm]{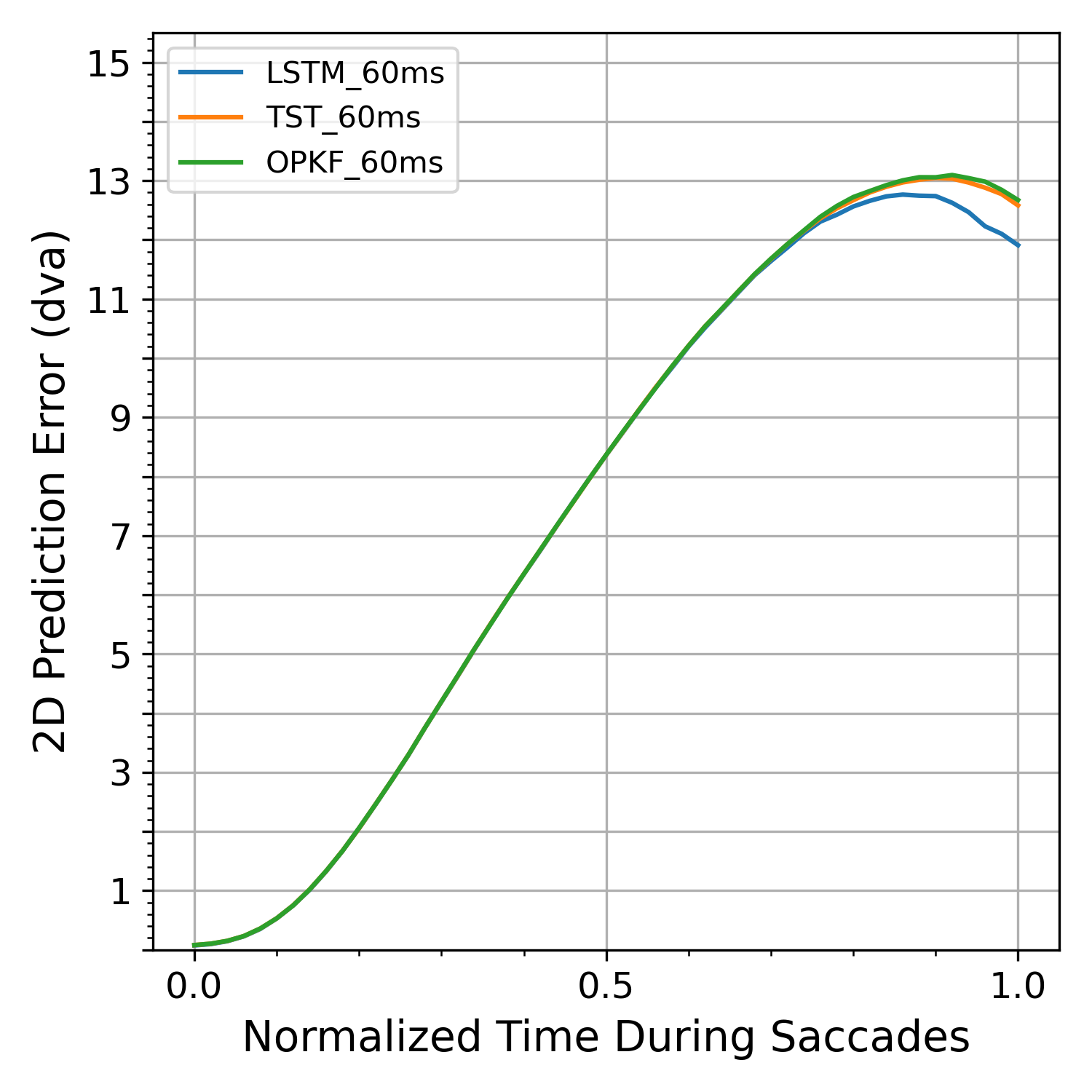}
        \caption{60 PI}
        \label{fig:app_60_sac_stat}
    \end{subfigure}
    \caption{Gaze Prediction Errors During Saccade Progression Across PIs}
    \label{fig:app_sac_stat}
\end{figure*}

In Fig. \ref{fig:app_sac_stat}, we present the prediction error as a function of normalized time during the progression of saccades with amplitudes in the range [10, 20]. The same analysis is presented for three prediction intervals. It is clear that as the prediction interval increases, the prediction error of saccades also increases. 

\begin{figure*}
    \begin{subfigure}{0.32\textwidth}
        \includegraphics[width=\textwidth, height=3.9cm]{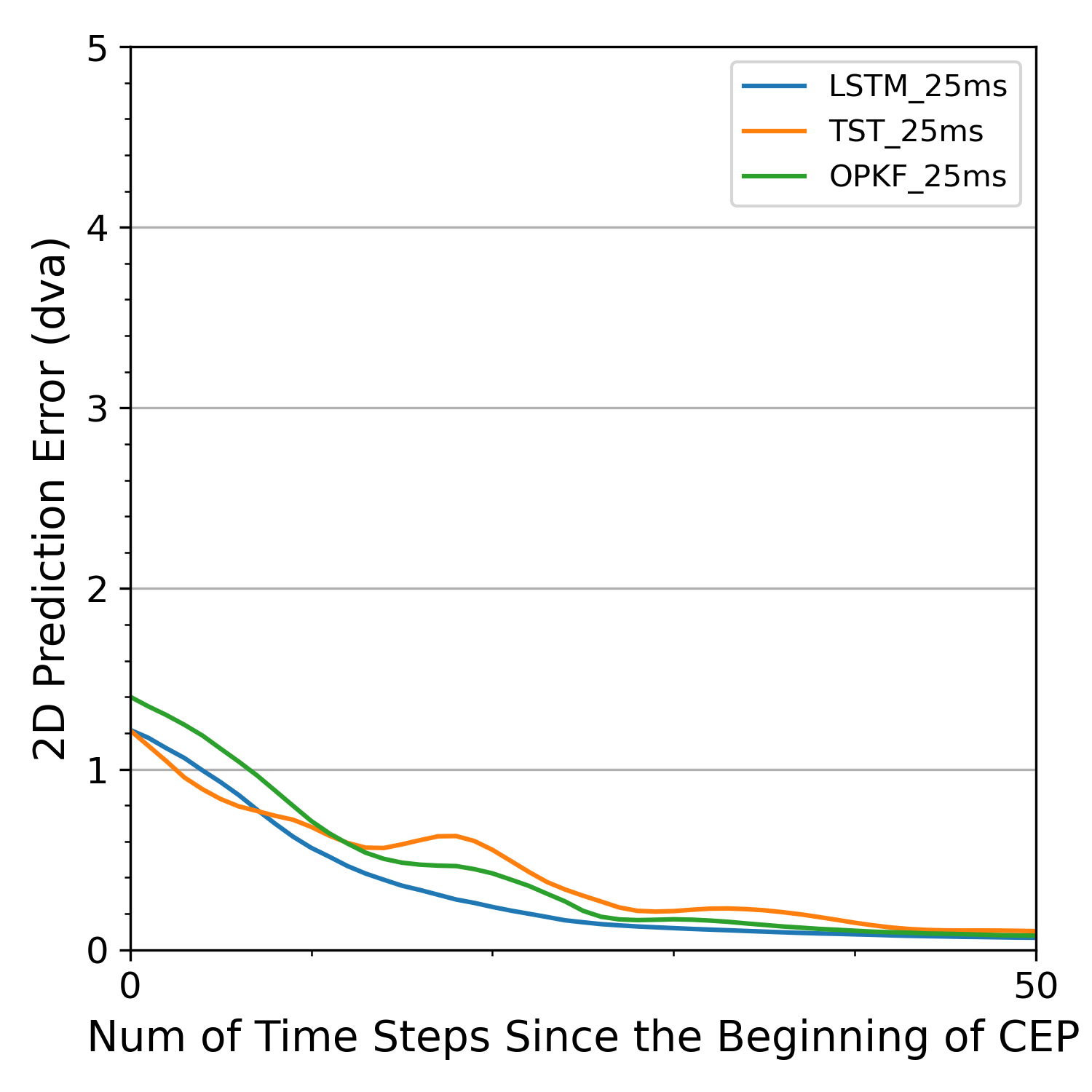}
        \caption{25 PI}
        \label{fig:app_cep25}
    \end{subfigure}
    \hfill
    \begin{subfigure}{0.32\textwidth}
        \includegraphics[width=\textwidth, height=3.9cm]{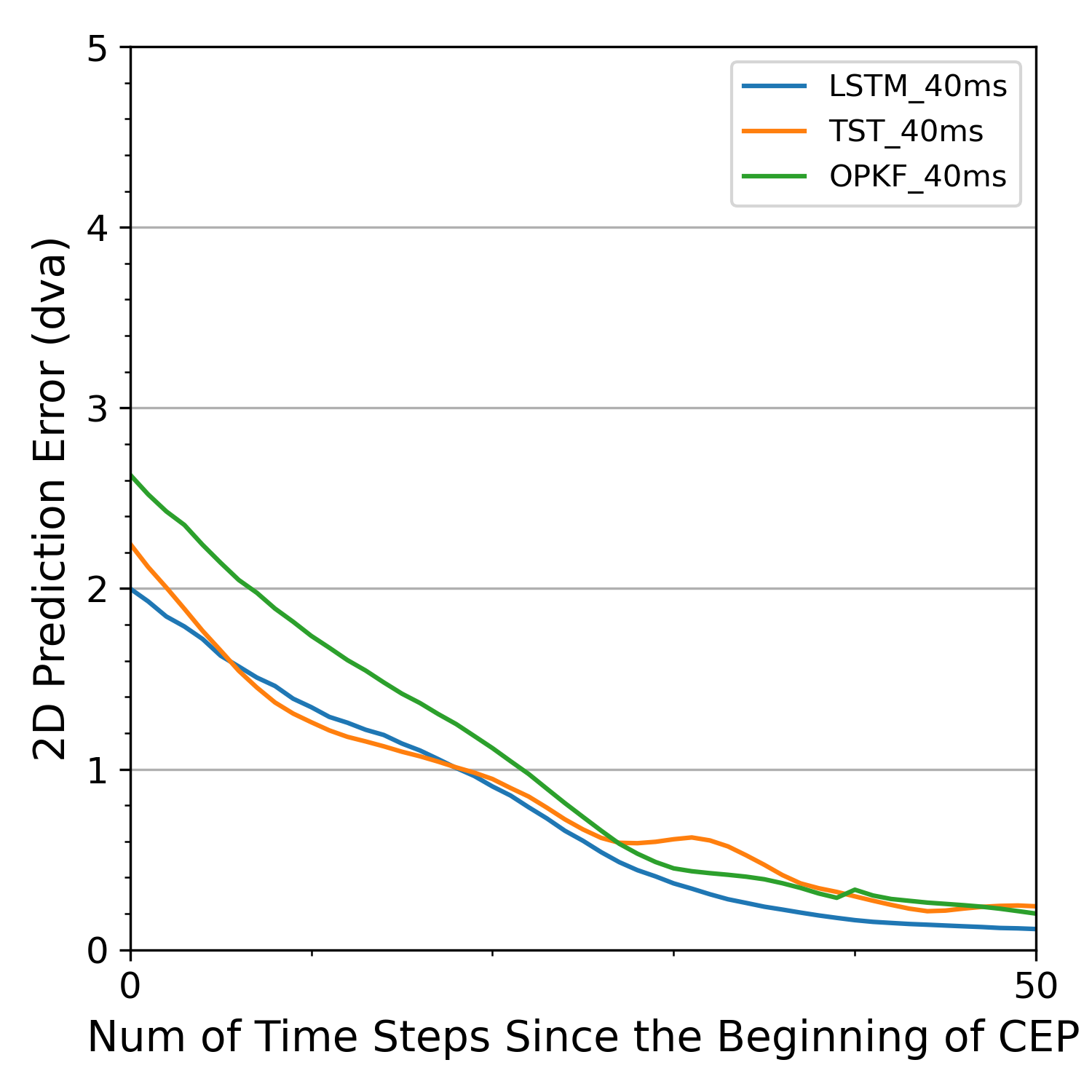}
        \caption{40 PI}
        \label{fig:app_cep40}
    \end{subfigure}
    \hfill
    \begin{subfigure}{0.32\textwidth}
        \includegraphics[width=\textwidth, height=3.9cm]{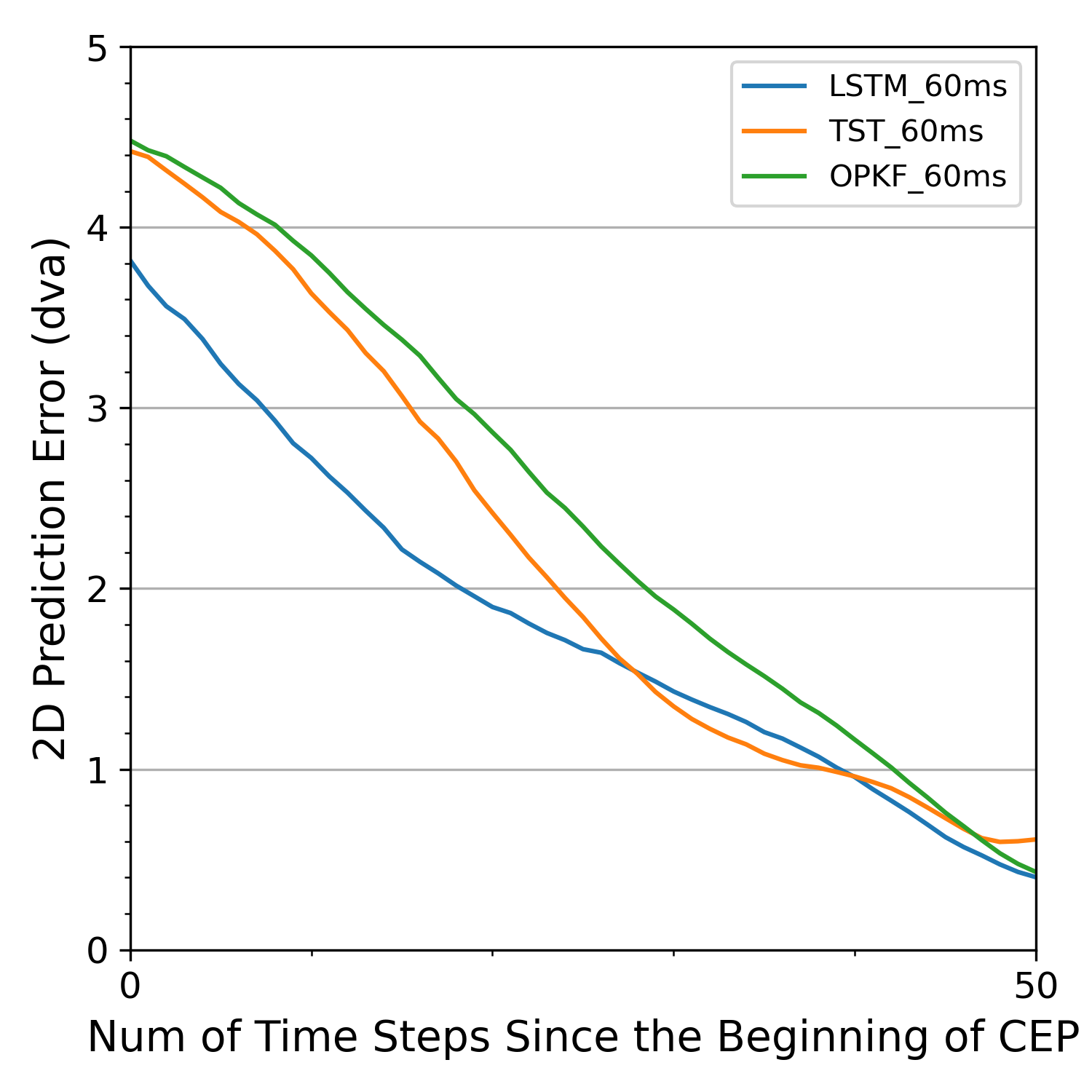}
        \caption{60 PI}
        \label{fig:app_cep60}
    \end{subfigure}
    \caption{Gaze Prediction Errors During First 50 ms of CEP Across PIs}
    \label{fig:app_cep_20_40_60}
\end{figure*}

In Fig. \ref{fig:app_cep_20_40_60}, we present prediction error as a function of time through the first half of the CEP interval. Once again, longer PIs are associated with larger errors.

\section{Evaluating Model Differences in Subject Profiles a Function of PI}
We performed a reliability analysis using Kendall's Coefficient of Concordance (KCC) \cite{KCC}. This analysis would most typically compare different raters to get inter-rater reliability or compare different time points to measure temporal persistence. KCC ranges from 0 to 1.0, with 1.0 indicating perfect agreement. In our case, each model is considered as a different rater, and what we get is a measure of inter-model agreement. 

\begin{figure*}
    \caption{Subject Profiles in Gaze Prediction Error Per Fixation Across PIs}
    
    \begin{subfigure}{0.32\textwidth}
        \includegraphics[width=\textwidth]{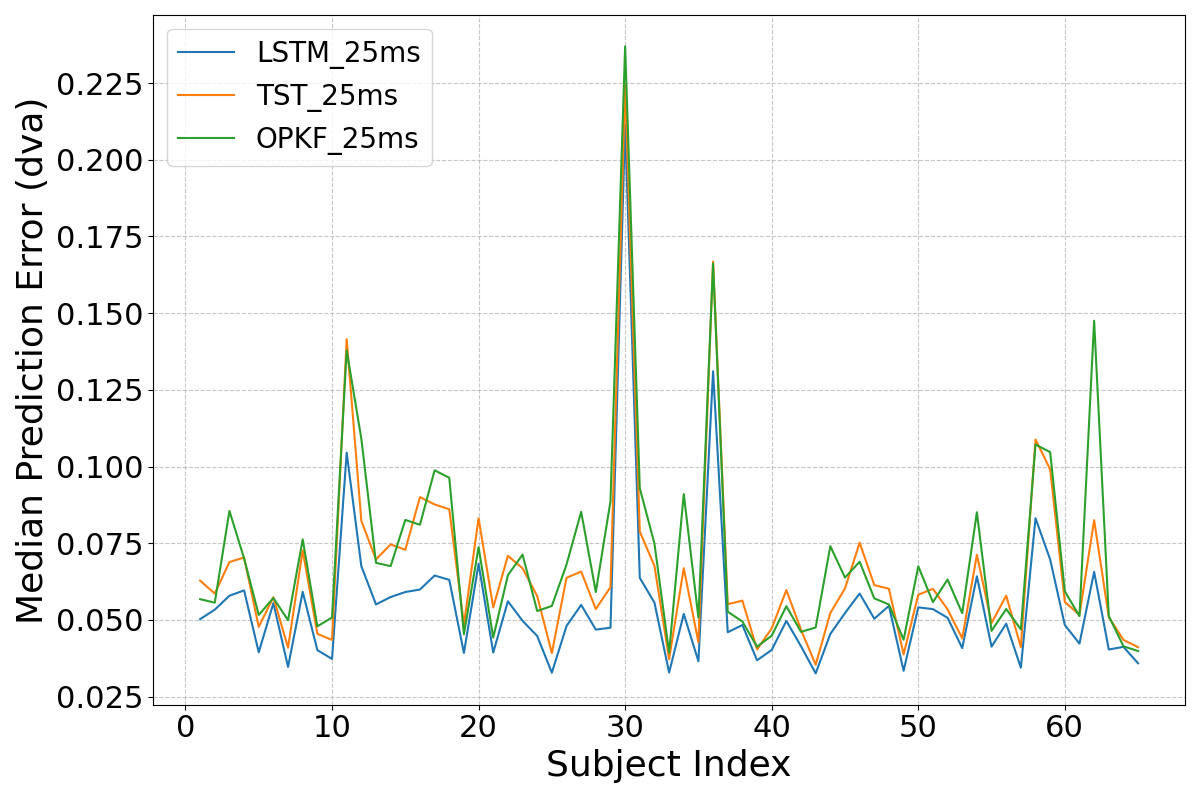}
        \caption{PI 25: KCC=0.944}
        \label{fig:fix_p50_25}
    \end{subfigure}
    \hfill
    \begin{subfigure}{0.32\textwidth}
        \includegraphics[width=\textwidth]{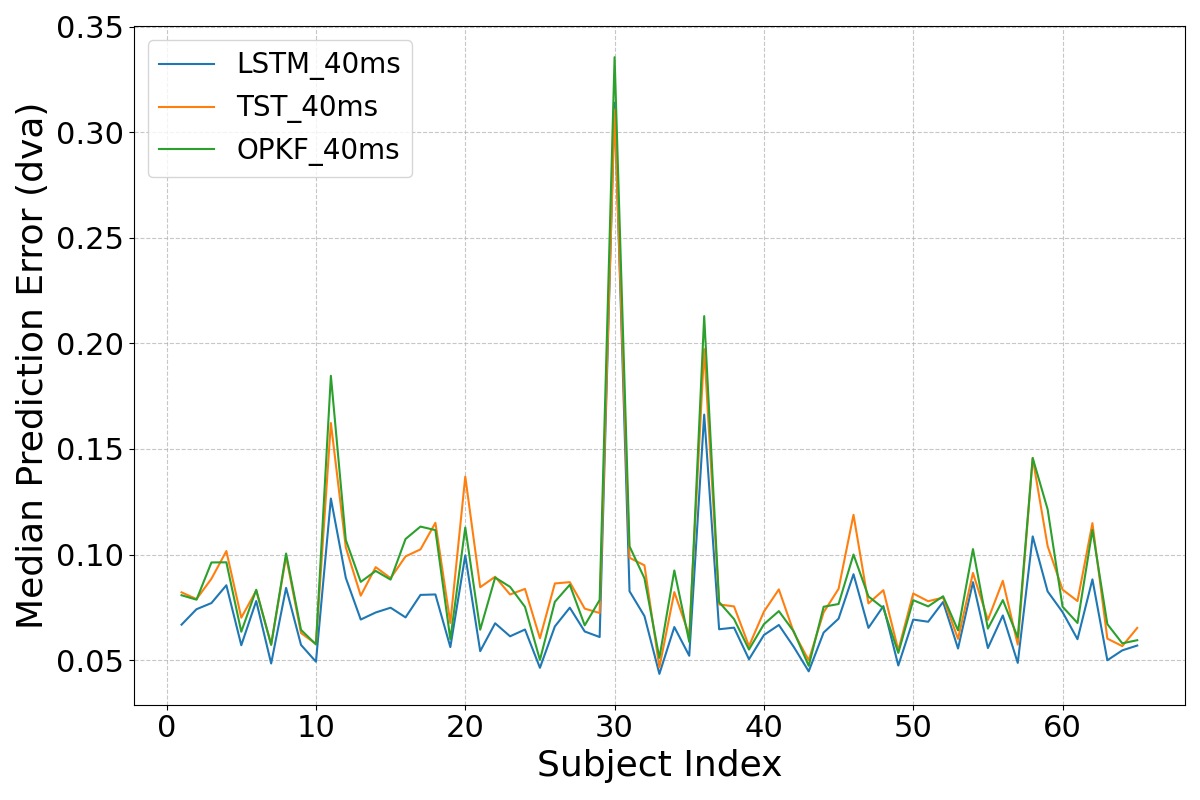}
        \caption{PI 40: KCC=0.949}
        \label{fig:fix_p50_40}
    \end{subfigure}
    \hfill
    \begin{subfigure}{0.32\textwidth}
        \includegraphics[width=\textwidth]{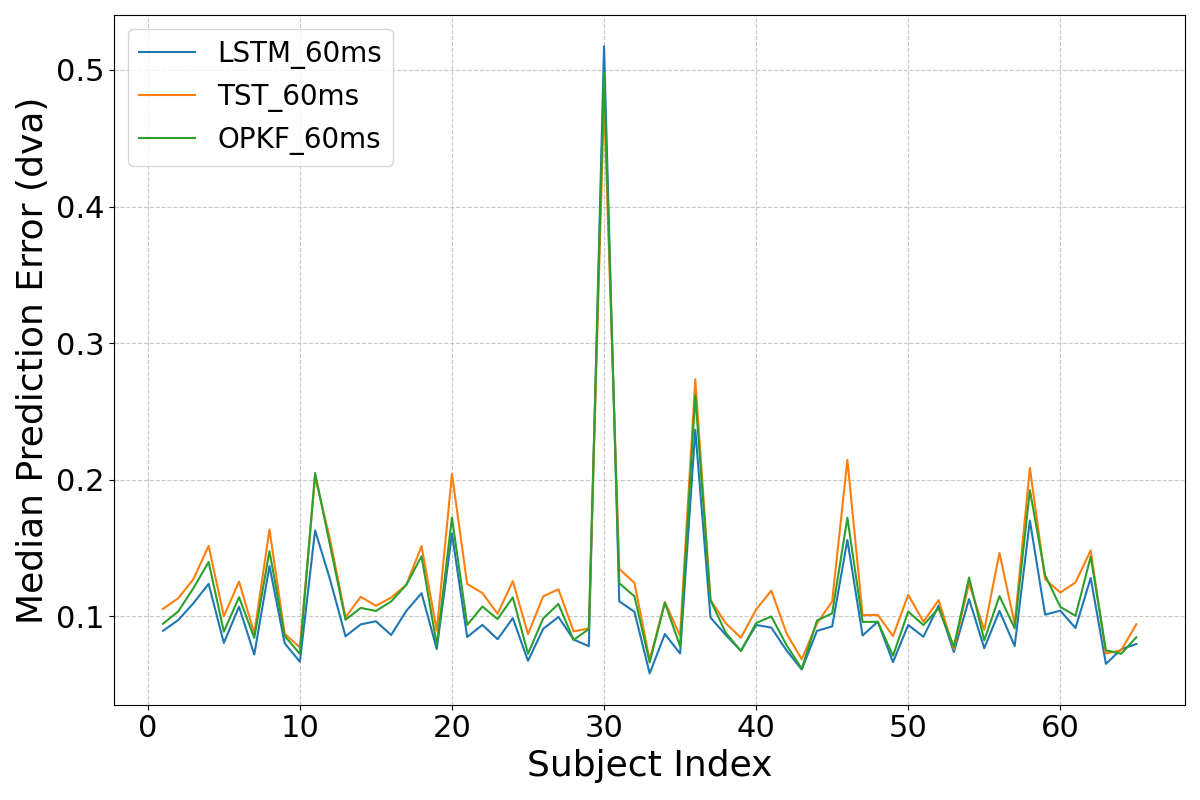}
        \caption{PI 60: KCC=0.97}
        \label{fig:fix_p50_60}
    \end{subfigure}
    
    \label{fig:fix_p50_25_40_60}
\end{figure*}

\begin{figure*}
    \caption{Subject Profiles in Gaze Prediction Error Per Large Saccades Across PIs}
    
    \begin{subfigure}{0.32\textwidth}
        \includegraphics[width=\textwidth]{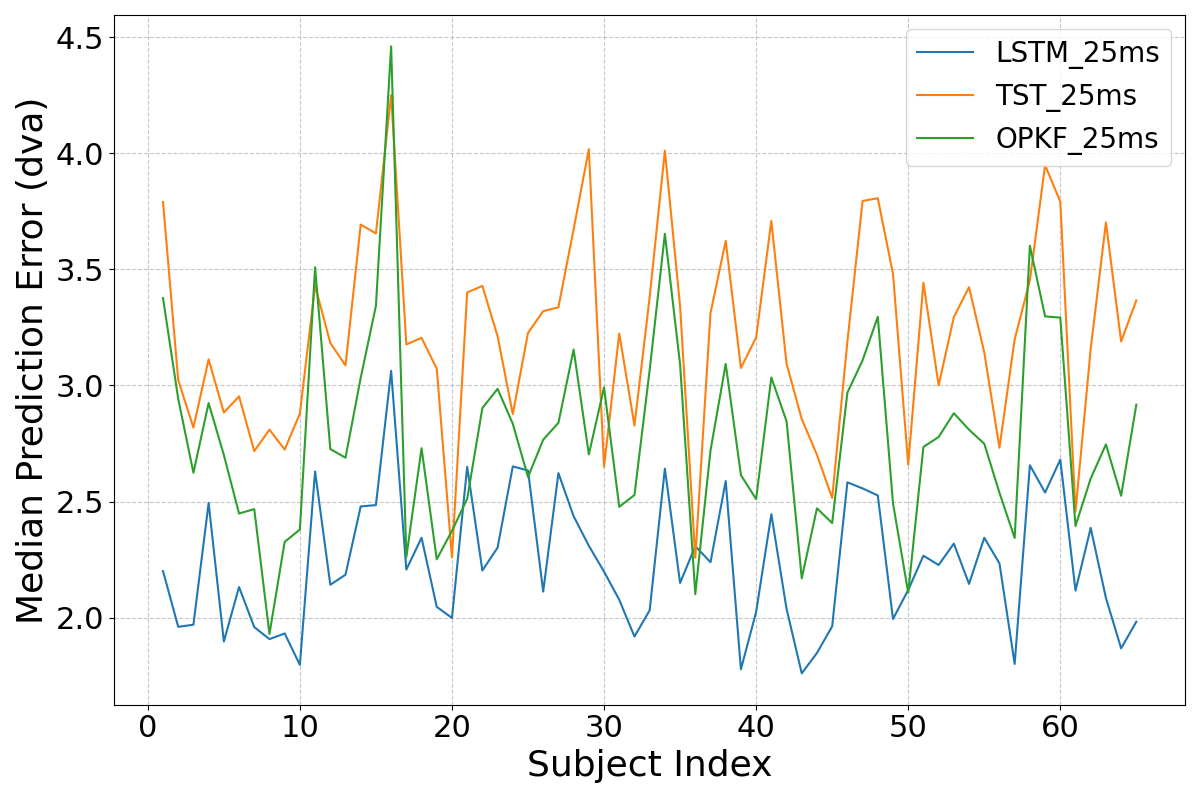}
        \caption{PI 25: KCC=0.763}
        \label{fig:lr_p50_25}
    \end{subfigure}
    \hfill
    \begin{subfigure}{0.32\textwidth}
        \includegraphics[width=\textwidth]{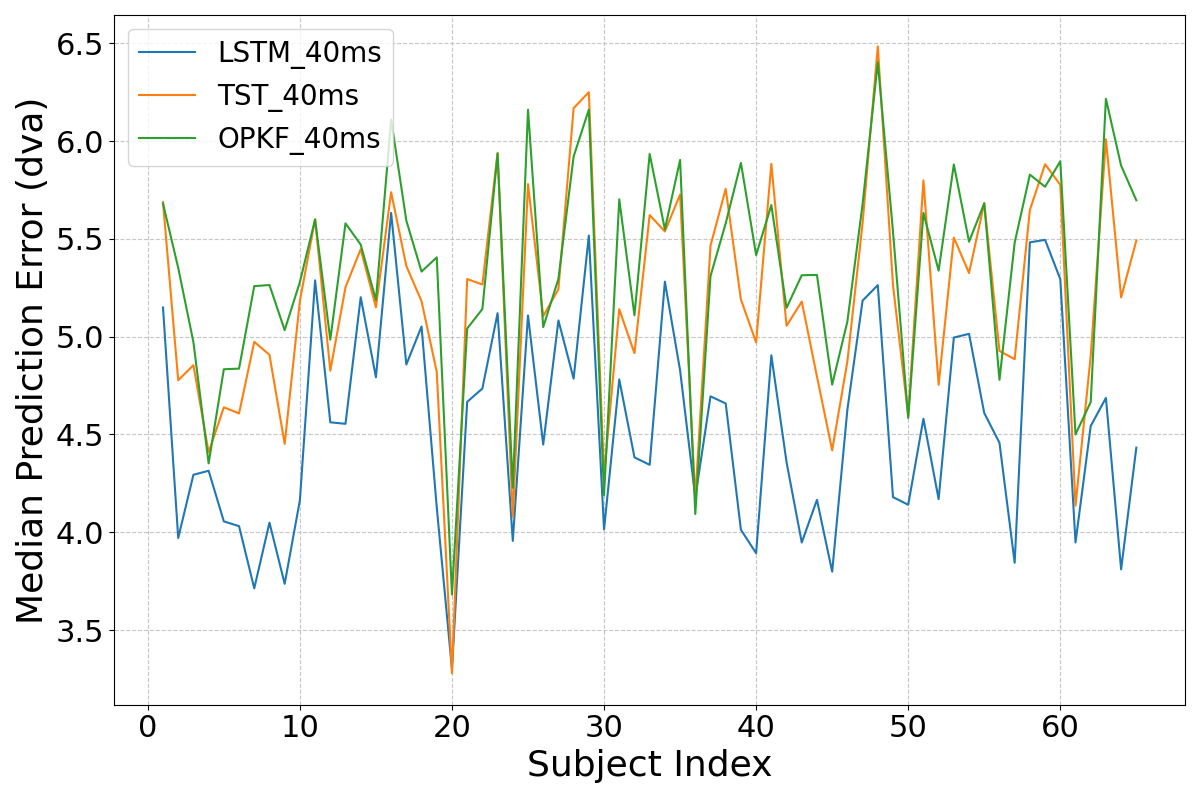}
        \caption{PI 40: KCC=0.828}
        \label{fig:lr_p50_40}
    \end{subfigure}
    \hfill
    \begin{subfigure}{0.32\textwidth}
        \includegraphics[width=\textwidth]{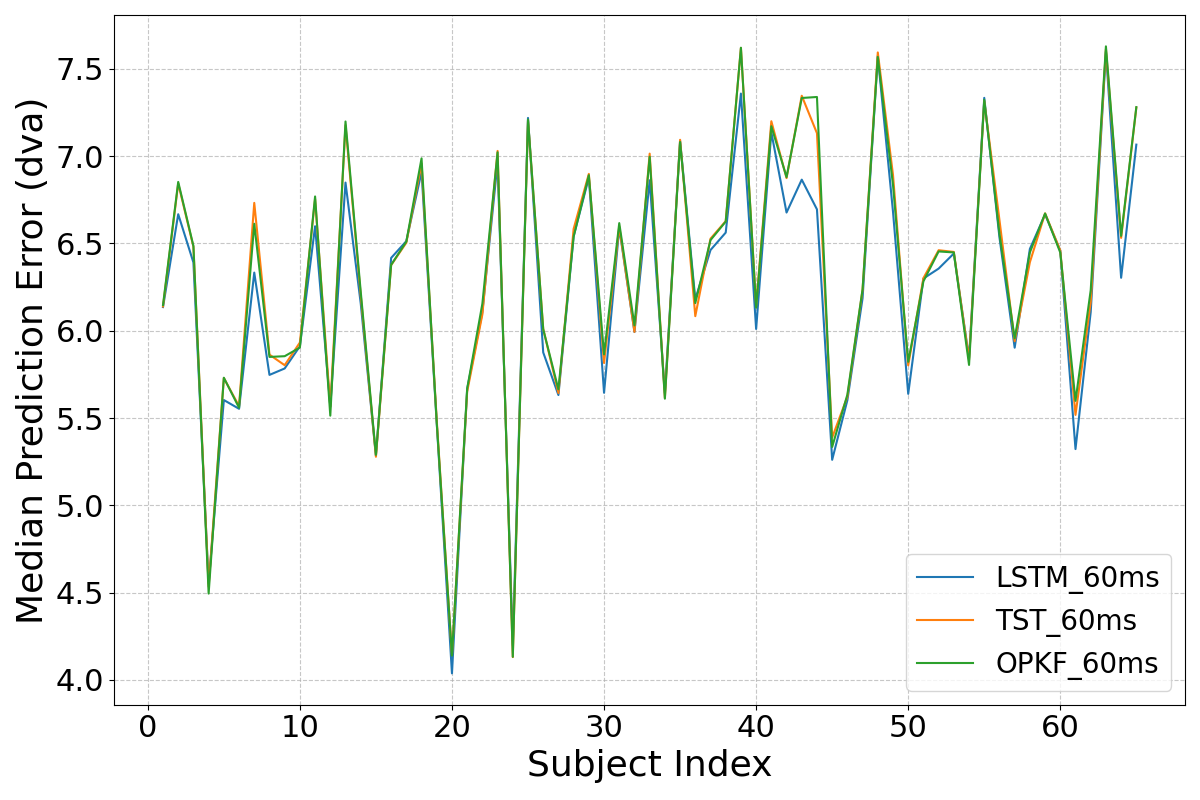}
        \caption{PI 60: KCC=0.993}
        \label{fig:lr_p50_60}
    \end{subfigure}
    
    \label{fig:lr_p50_25_40_60}
\end{figure*}

\begin{figure*}
    \caption{Subject Profiles in Gaze Prediction Error Per Small Saccades Across PIs}
    
    \begin{subfigure}{0.32\textwidth}
        \includegraphics[width=\textwidth]{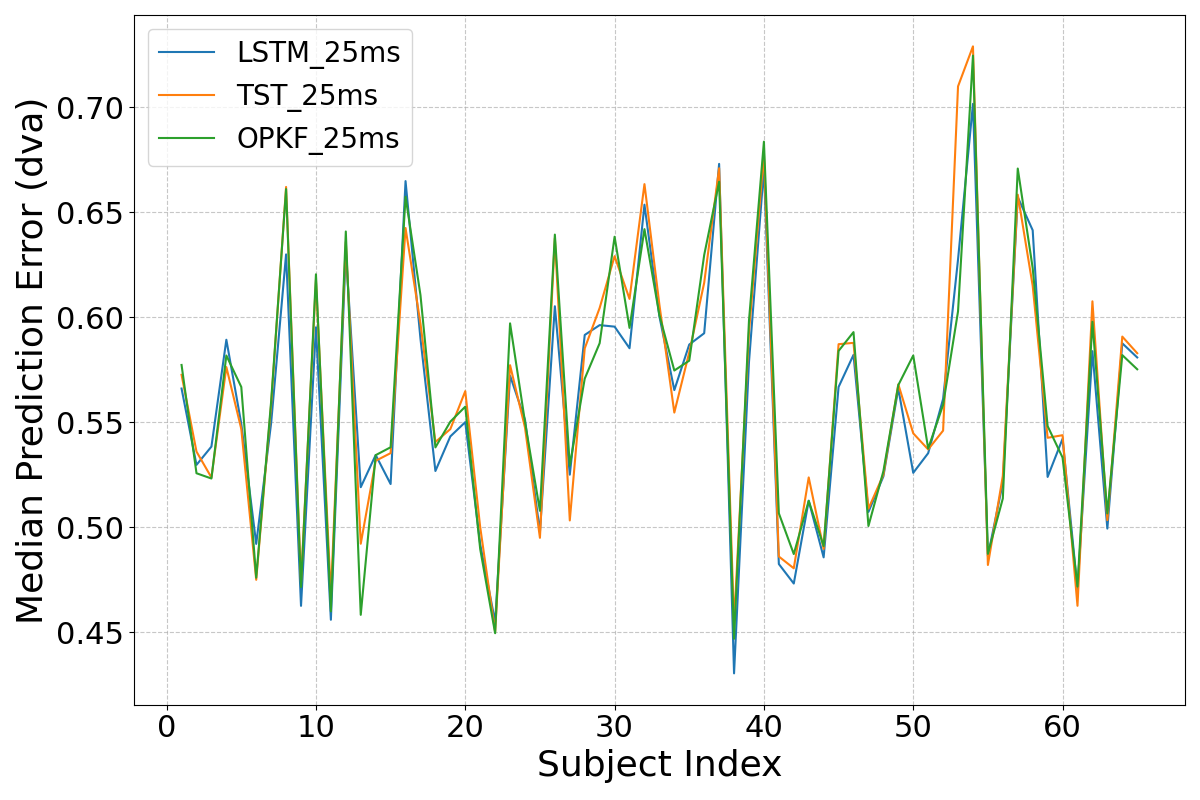}
        \caption{PI 25: KCC=0.982}
        \label{fig:sm_p50_25}
    \end{subfigure}
    \hfill
    \begin{subfigure}{0.32\textwidth}
        \includegraphics[width=\textwidth]{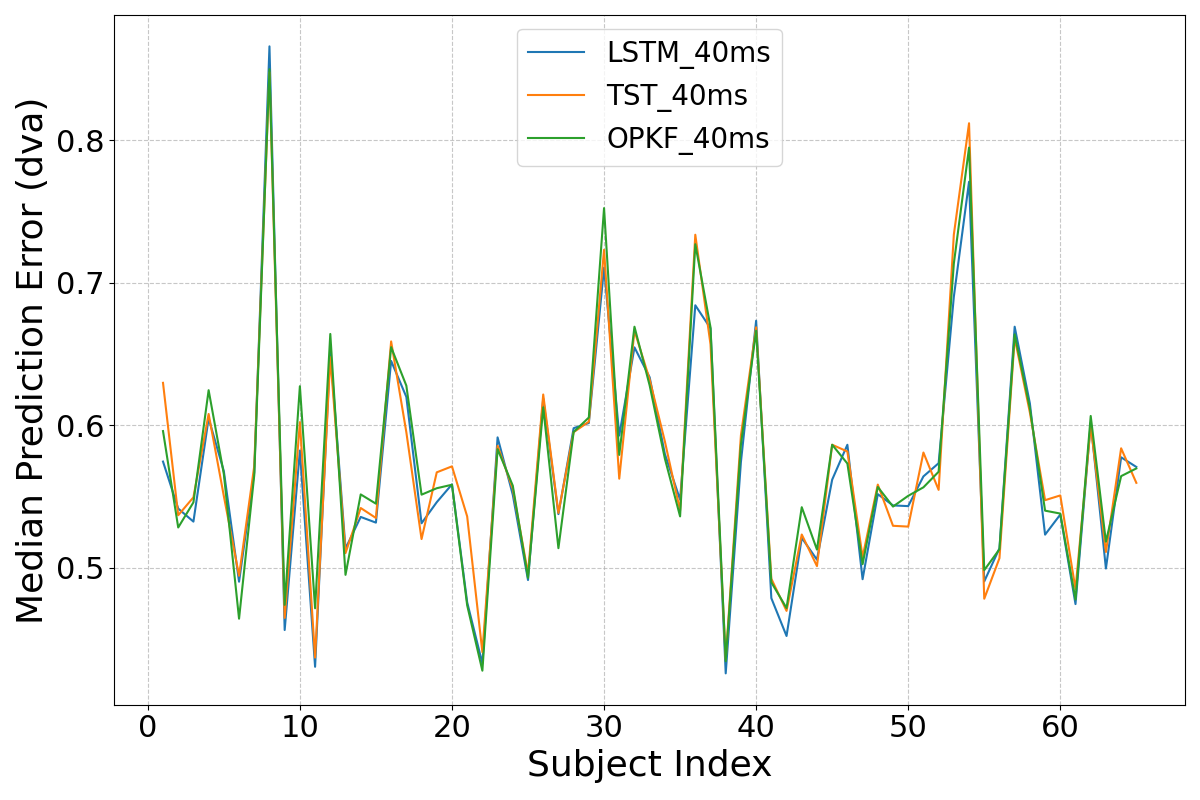}
        \caption{PI 40: KCC=0.982}
        \label{fig:sm_p50_40}
    \end{subfigure}
    \hfill
    \begin{subfigure}{0.32\textwidth}
        \includegraphics[width=\textwidth]{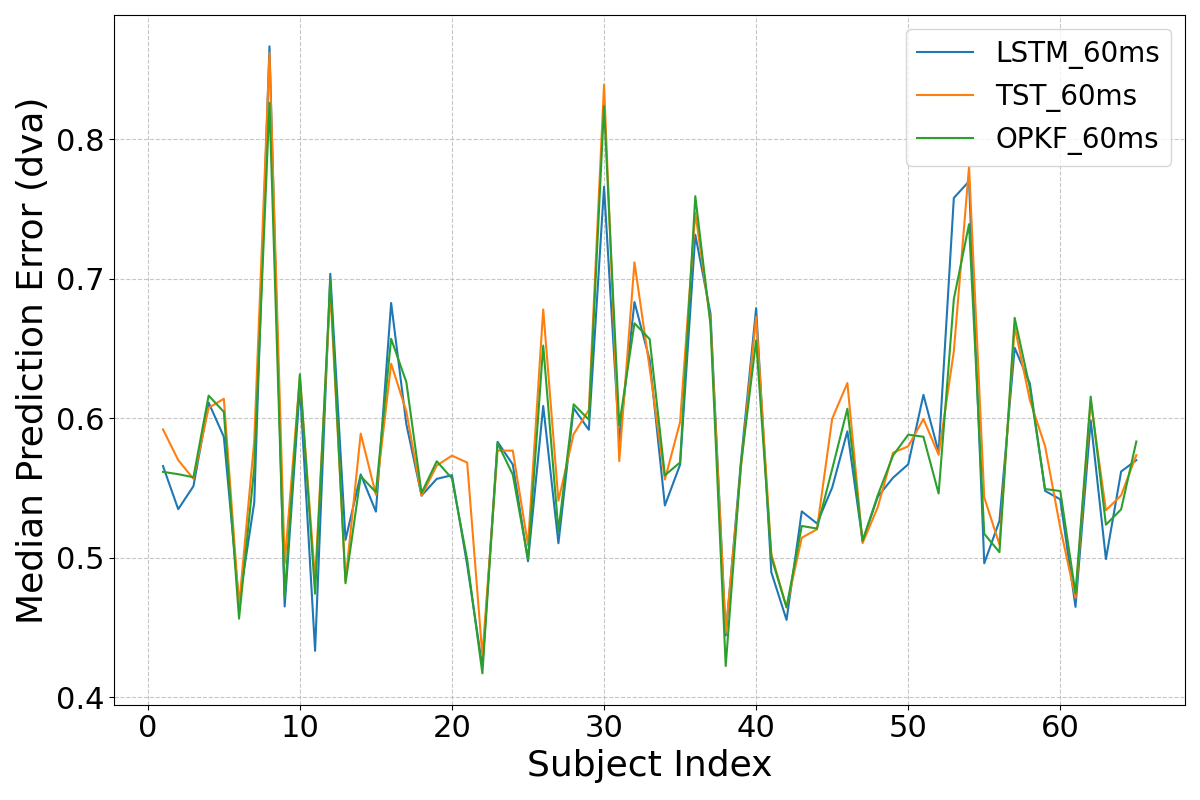}
        \caption{PI 40: KCC=0.968}
        \label{fig:sm_p50_60}
    \end{subfigure}
    
    \label{fig:sm_p50_25_40_60}
\end{figure*}

In Fig. \ref{fig:fix_p50_25_40_60}, the subject profiles for all models are presented for increasing PIs during fixation. We also provide the KCC values representing model performance similarity for each PI interval in the caption of each subfigure. As PI increases, we can see that the model performance becomes more similar. It is also evident for large saccades Fig. \ref{fig:lr_p50_25_40_60}, but not for small saccades (see in Fig. \ref{fig:sm_p50_25_40_60}).

For fixations Fig. \ref{fig:fix_p50_25_40_60}, the KCC, based on median performance, ranged from $0.944$ to $0.97$, indicating a high degree of inter-model agreement. It is also true for small saccades (see Fig. \ref{fig:sm_p50_25_40_60}) (KCC from $0.968$ to $0.982$). For large saccades (see Fig. \ref{fig:lr_p50_25_40_60}), the agreement was slightly lower at PI equal to 25 (KCC=$0.763$) but still substantial. It was larger for the PI equal to 60 (KCC=$0.993$).

\section{Evaluating Inter-Model Agreement across PIs as a Function of Eye-Movement Type}
Table \ref{tab:mods_kccs_combined} clearly shows that model agreement is higher for fixations and small saccades and lower for large saccades.

\begin{table}
    \centering
    \begin{tabular}{|l|c|c|c|}
        \hline
        \multirow{2}{*}{Model} & \multicolumn{3}{c|}{Eye-Movement Type $P_{50}$} \\ \cline{2-4} 
        & Fixations & Large Saccades & Small Saccades \\ \hline
        LSTM  & 0.952 & 0.584 & 0.968 \\ \hline
        TST   & 0.925 & 0.715 & 0.943 \\ \hline
        OPKF  & 0.934 & 0.621 & 0.962 \\ \hline
    \end{tabular}
    \caption{Level of Agreement (KCC) as a Function of Model Type and Eye-Movement Type}
    \label{tab:mods_kccs_combined}
\end{table}

\end{document}